\documentclass{article}
\usepackage{ijcai25}

\usepackage{times}
\usepackage{soul}
\usepackage{url}
\usepackage[hidelinks]{hyperref}
\usepackage[utf8]{inputenc}
\usepackage[small]{caption}
\usepackage{graphicx}
\usepackage{amsmath}
\usepackage{amsthm}
\usepackage{booktabs}
\usepackage{algorithm}
\usepackage{algorithmicx}
\usepackage{algpseudocode}
\usepackage{amsmath}
\usepackage{hyperref}

\usepackage{makecell}

\usepackage{listings}
\usepackage{xcolor}

\usepackage{multirow}

\usepackage{amssymb}
\usepackage{pgfplots}

\definecolor{mypink}{rgb}{0.95, 0.5, 0.7}
\definecolor{mycolor_1}{rgb}{0.91, 0.99, 0.91}
\definecolor{mycolor_2}{rgb}{0.12, 0.46, 0.87}
\definecolor{mycolor_3}{rgb}{0.77, 0.37, 0.06}

\usepackage{hyperref}

\hypersetup{
    colorlinks=true,
    linkcolor=black,
    filecolor=black,      
    urlcolor=black,
    citecolor=black
}

\title{\includegraphics[width=0.06\linewidth]{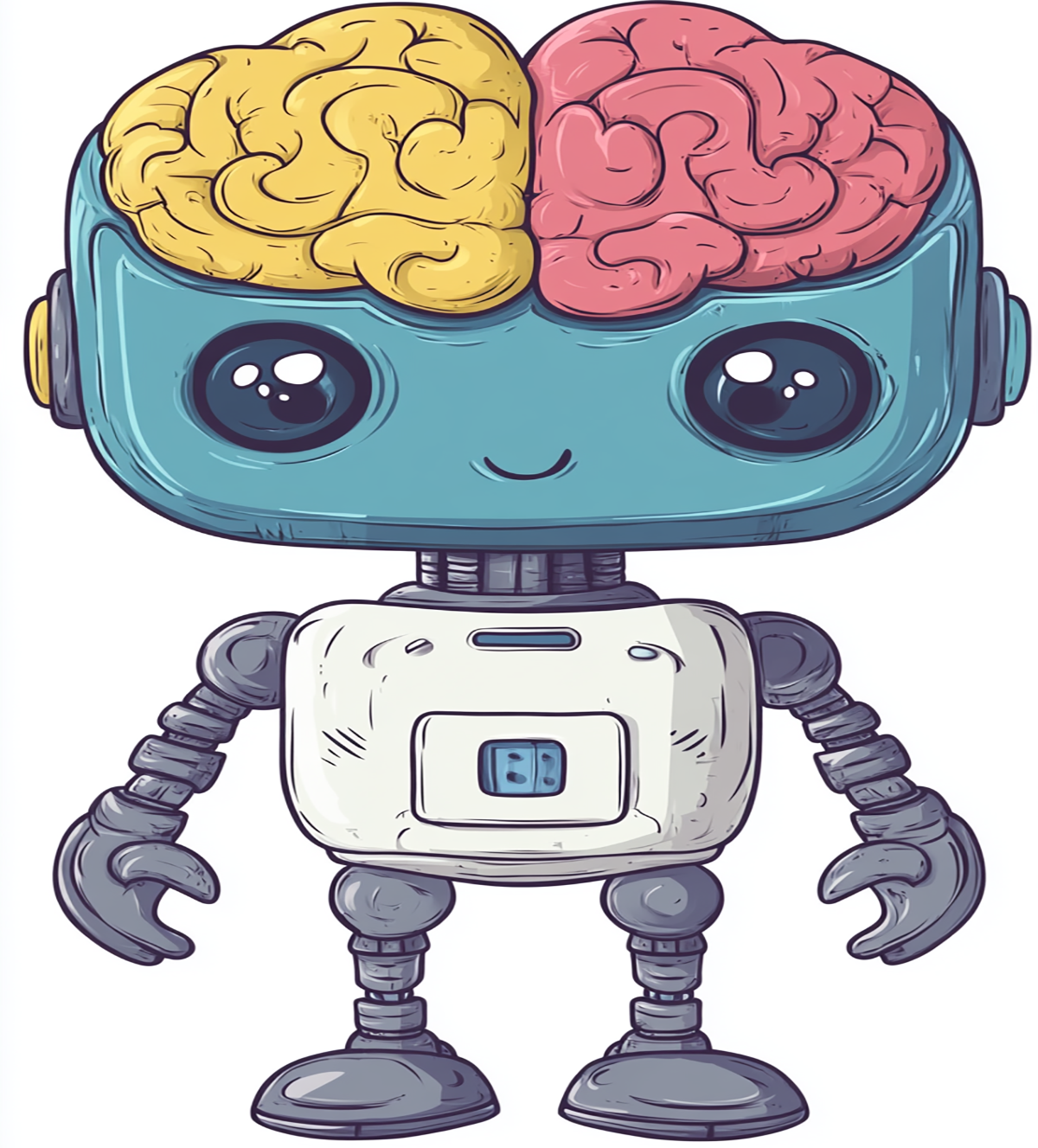} Cogito, ergo sum: A Neurobiologically-Inspired Cognition-Memory-Growth System for Code Generation}

\author{
Yanlong Li$^1$
\and
Jindong Li$^2$\and
Qi Wang$^3$\footnote{Corresponding author.}\and
Menglin Yang$^{2}$\and
He Kong$^{3}$\and
Shengsheng Wang$^{1}$*\\
\affiliations
$^1$College of Computer Science and Technology, Jilin University\\
$^2$The Hong Kong University of Science and Technology (Guangzhou)\\
$^3$School of Artificial Intelligence, Jilin University\\
\emails
yanlong23@mails.jlu.edu.cn,
jindong.li.mail@gmail.com,
qiwang@jlu.edu.cn,
menglin.yang@outlook.com,
konghe19@mails.jlu.edu.cn,
wss@jlu.edu.cn
}

\begin{document}

\maketitle

\begin{abstract}

Large language models-based Multi-Agent Systems (MAS) have demonstrated promising performance for enhancing the efficiency and accuracy of code generation tasks. However, most existing methods follow a conventional sequence of planning, coding, and debugging, which contradicts the growth-driven nature of human learning process. Additionally, the frequent information interaction between multiple agents inevitably involves high computational costs. In this paper, we propose \textbf{Cogito}, a neurobiologically-inspired multi-agent framework to enhance the problem-solving capabilities in code generation tasks with lower cost. Specifically, \textbf{Cogito} adopts a reverse sequence: it first undergoes debugging, then coding, and finally planning. This approach mimics human learning and development, where knowledge is acquired progressively. Accordingly, a hippocampus-like memory module with different functions is designed to work with the pipeline to provide quick retrieval in similar tasks. Through this growth-based learning model, \textbf{Cogito} accumulates knowledge and cognitive skills at each stage, ultimately forming a Super-Role—an all-capable agent to perform the code generation task. Extensive experiments against representative baselines demonstrate the superior performance and efficiency of \textbf{Cogito}. The code is publicly available at \url{https://github.com/doc0318/Cogito}.
\end{abstract}

\begin{figure}[t]
    \centering
    \includegraphics[width=0.99\linewidth]{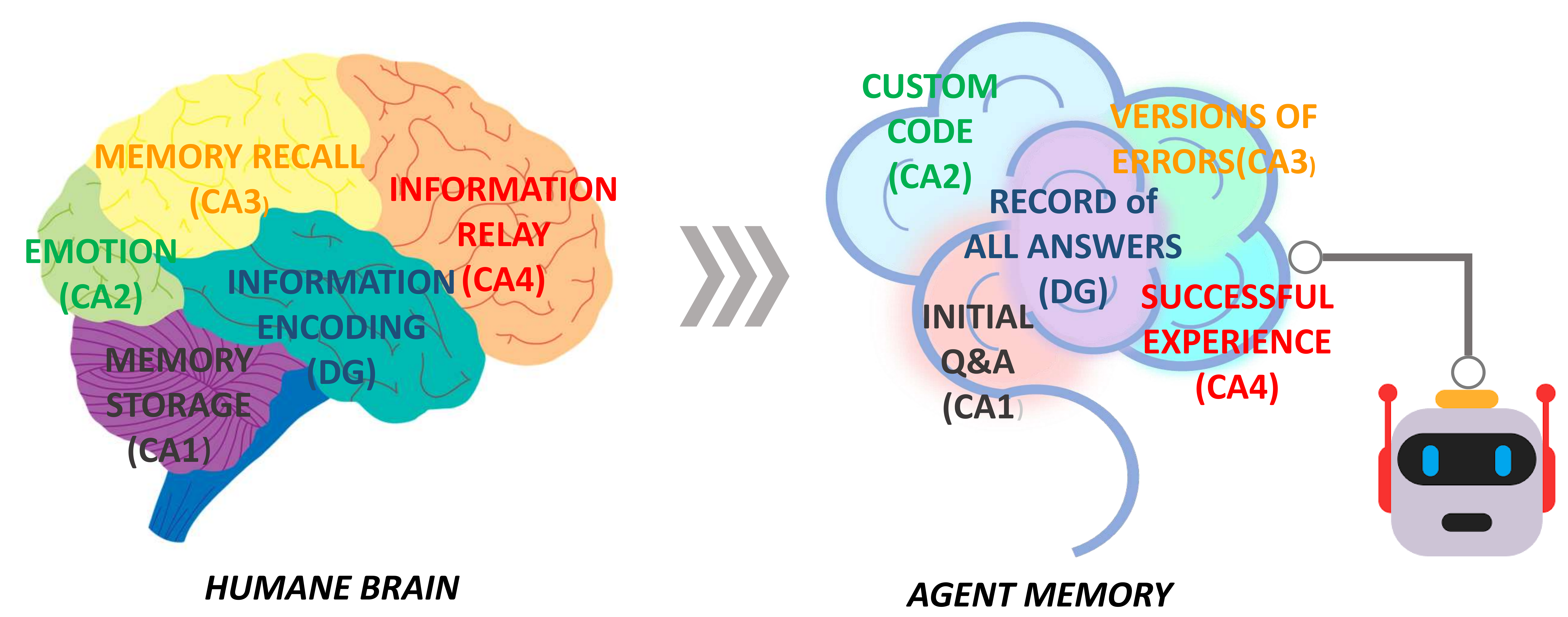}
    \includegraphics[width=0.86\linewidth]{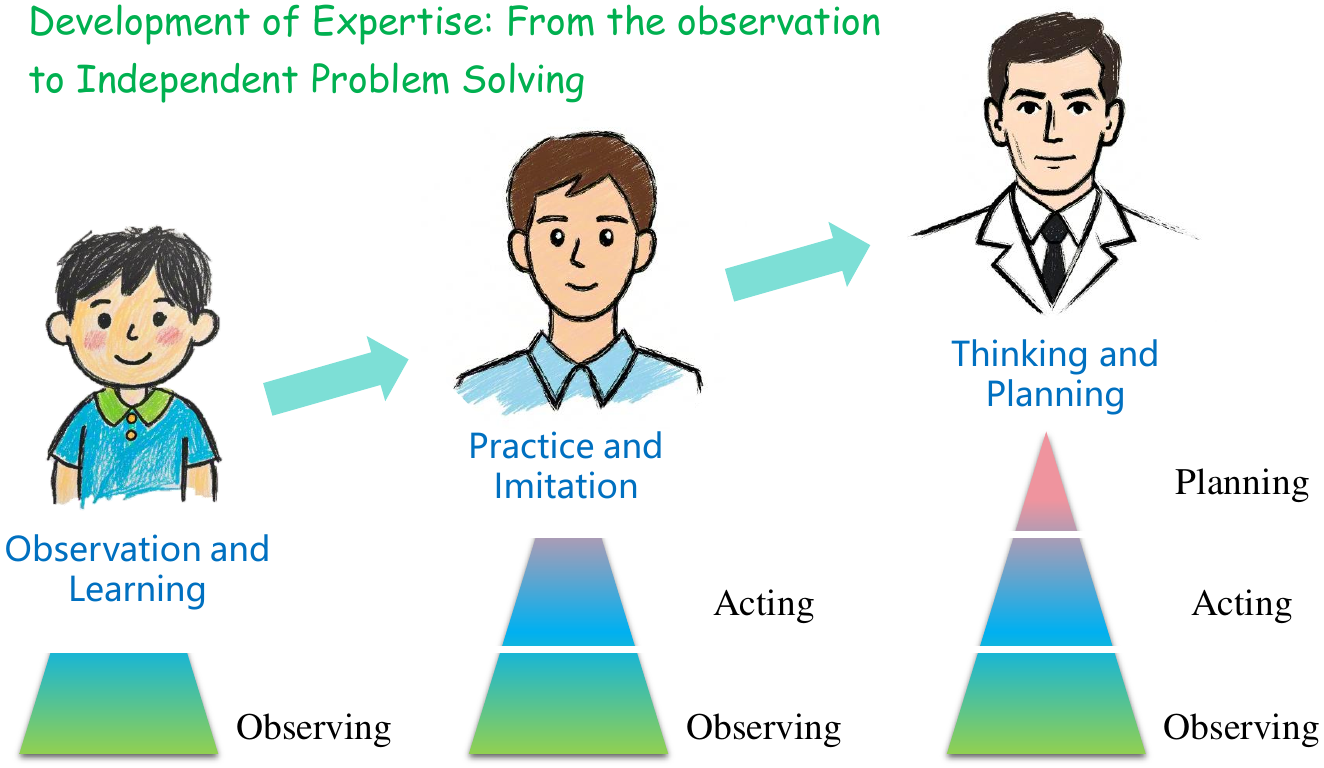}
    \caption{The intuitions behind this work. (Top): brain’s different regions are dedicated to distinct functions and tasks. Inspired by this functional specialization, we design an agent with distinct roles that evolve through stages. (Bottom): the growth trajectory of an individual, progressing from observation and learning in childhood, to practice and imitation in young adulthood, and finally to independent problem-solving and planning in the expert stage.}
    \label{fig:fig_1}
\end{figure}

\section{Introduction}
Large language models (LLMs) have demonstrated human-like intelligence in tasks such as code generation~\cite{2022_arXiv_PaLM}, testing~\cite{2024_arXiv_LLMBased}, and debugging~\cite{2023_arXiv_Conversational}. Recent studies show the effectiveness of using multiple agents for collaborative tasks, achieving superior performance over single-agent~\cite{2024_ACL_MapCoder,2024_arXiv_CodePori}. Such advancements not only elevate the models' proficiency in automating segments of the development lifecycle but also encompass refining the models' adeptness at deciphering intricate problems and performing sophisticated logical reasoning. As a result, code generation has garnered considerable attention from academic researchers, industry professionals, as well as institutions like OpenAI~\footnote{\url{https://api.semanticscholar.org/CorpusID:257532815}} and Meta AI~\footnote{\url{https://api.semanticscholar.org/CorpusID:271571434}}.

While existing works have shown promising results, these methods typically follow a standardized programming workflow of human programmers, e.g., first planning, then coding, finally debuging~\cite{2024_ACL_MapCoder,2024_ICLR_MetaGPT}. However, a deeper understanding, grounded in practical experience, is essential to appreciate the underlying principles and navigate the transition from learning to doing and ultimately to directing. As shown in Figure~\ref{fig:fig_1}, individuals often begin with observation and learning, gradually gaining expertise to practice and imitation, and finally thinking and planning. Such a process renders a reverse process from existing code generation pipelines. Also, multi-agent frameworks inherently require agent communication. As the number of agents increases, multi-agent collaborative communication inevitably raises the overall computational and operational expenses~\cite{2023_arXiv_AgentCoder}.

Additionally, existing methods for external storage either store the complete responses directly~\cite{2024_arXiv_XUATCopilot} or summarize all the content before storing~\cite{2023_arXiv_Experiential}. These approaches either require excessive storage space or result in content loss due to summarization. As shown in Figure \ref{fig:fig_1}, human brain excels in complex tasks through a network of specialized regions responsible for functions like sensory processing, decision-making, and memory storage~\cite{2024_Journal_of_NeuroScience_Functional_Specialization_in_the_Human_Brain_Estimated_by_Intrinsic_Hemispheric_Interaction}. This specialization promotes efficient cognitive performance as regions collaborate to process and integrate information~\cite{2024_Nature_Reviews_Neuroscience_Structure_Function_Coupling_in_Macroscale_Human_Brain_Networks}. The hippocampus, vital for memory storage and retrieval, plays a key role in problem-solving and decision-making~\cite{1986_APA_The_Hippocampal_Memory_Indexing_Theory}.

Drawing inspiration from these principles, we introduce a neurobiologically-inspired multi-agent framework, \textbf{Cogito}, which aims to enhance the problem-solving capabilities in code generation tasks while reducing costs. Different from existing works, \textbf{Cogito} employs a reverse sequence: starting with code analysis and correction as a Debugger, progressing to providing initial solutions as a Coder, and ultimately optimizing strategies as a Planner. This sequence mirrors the human learning and developmental process, where knowledge is acquired incrementally. Specifically, we enhance inter-agent communication by defining roles redundantly across agents, ensuring efficient information exchange to reduce communication costs. 
Accordingly, a hippocampus-like memory module with distinct functionalities is integrated into the pipeline, facilitating rapid retrieval in similar tasks. \textbf{Cogito} accumulates knowledge and cognitive skills at each stage, ultimately evolving into a super-agent for code generation. We conduct extensive experiments on eight widely used datasets, and the results demonstrate that Cogito can not only achieve state-of-the-art performance but also excel in resource consumption. 
\textbf{The key contributions are summarized as follows}:
\begin{itemize} 
    \item We present a unique growth-based learning approach \textbf{Cogito}, enabling a \textbf{Super-Role} to evolve reversely from a Debugger to a Coder and eventually to a Planner, enhancing its problem-solving capabilities with less token consumption through accumulated experience.
    \item We design a hippocampus-inspired memory that stores different content based on learning stages, where different parts of memory are interconnected to ensure the completeness of stored information. The design can support dynamic and adaptive programming workflows. 
    \item We conduct extensive experiments to validate \textbf{Cogito}'s efficiency on eight code generation tasks. The results show that \textbf{Cogito} reduces token consumption by up to 66.29\% and improves performance by an average of 12.2\% compared to MapCoder, using GPT-3.5-turbo and GPT-4.
\end{itemize}

\begin{figure*}[t]
    \centering
    \includegraphics[width=0.99\linewidth]{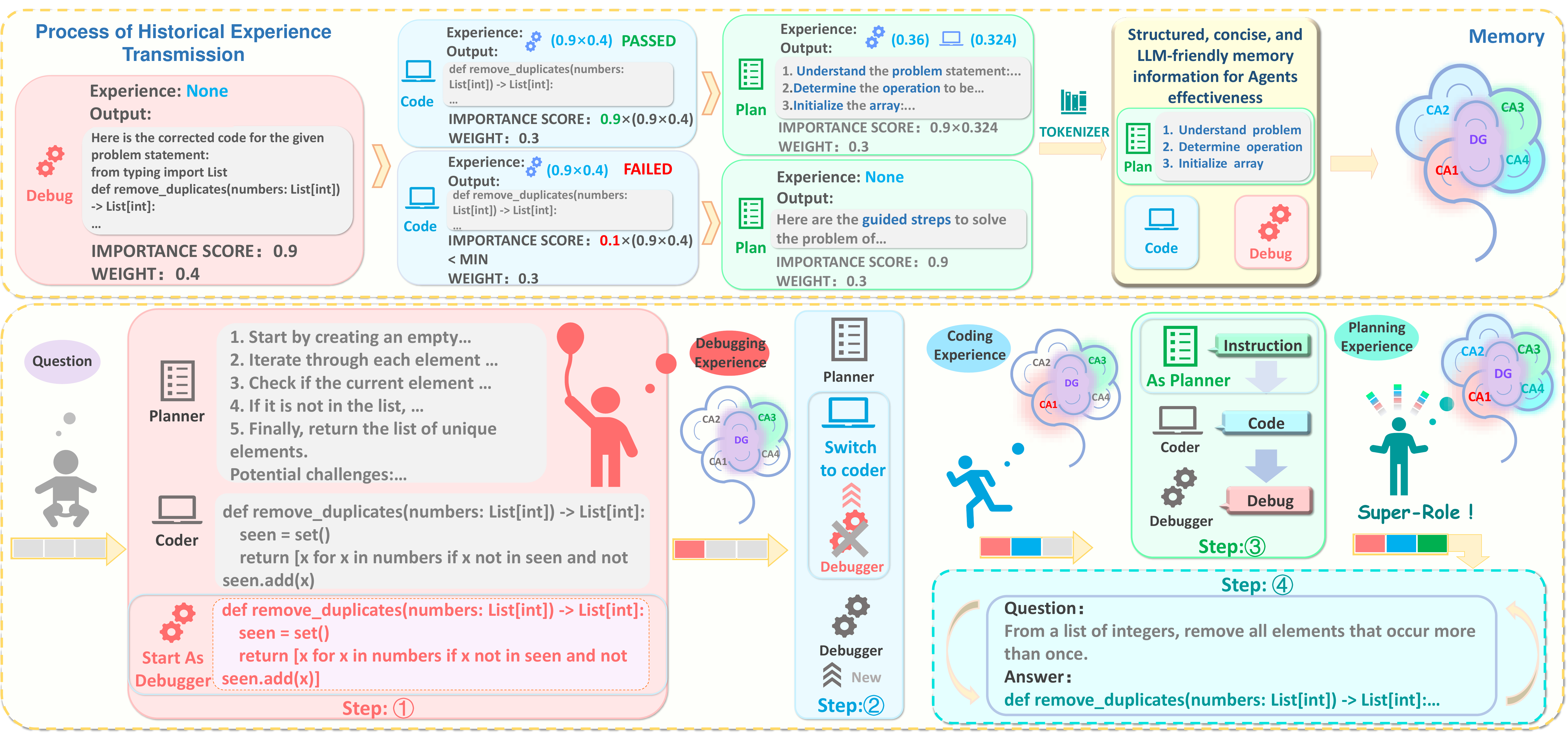}
    \caption{Overview of \textbf{Cogito}. The upper section illustrates the learning process of the \textbf{Super-Role} stored in the memory module. The lower section provides a detailed explanation of the process: initially, it assumes the role of the debugger within the group, followed by transitions to the coder and planner roles. After completing the learning cycle, the final answer is provided by the \textbf{Super-Role}.}
    \label{fig:framework}
\end{figure*}

\section{Related work}
\subsection{LLM Agents}
LLM-based agents normally consist of four core components: planning, memory, perception, and action. Planning and memory form the cognitive core, while perception and action enable interaction with the environment to achieve goals~\cite{2023_arXiv_The}. The planning component decomposes complex tasks into manageable subtasks and schedules their execution to achieve predefined objectives, while also incorporating the flexibility to adapt plans dynamically in response to external feedback. The memory component, on the other hand, stores historical actions and observations, enabling agents to draw on past experiences to refine decision-making processes and enhance task execution efficiency. This dual approach facilitates continuous learning and optimization, ensuring improved performance over time. Effective memory management is critical for system performance~\cite{2023_arXiv_A,2024_arXiv_A}. Due to the suitability of this setup for code generation problems, a large number of works have emerged in this field~\cite{2024_arXiv_Agents,2024_arXiv_Large}.

\subsection{Multi-Agent Collaboration for Software Development}
To effectively solve complex problems, tasks are divided into specialized roles, each handling a specific aspect of the process. This role-based division, combined with agent collaboration, boosts efficiency and enhances outcomes. The typical workflow includes task refinement, execution, result validation, and optimization~\cite{2024_Arxiv_4PlanningDriven,2024_arXiv_AutoCoder}. These stages ensure that each component is managed with focus, leading to smoother task execution and more reliable results. For instance, MetaGPT~\cite{2024_ICLR_MetaGPT} mimics standardized real-world collaboration procedures, incorporating five distinct roles. Similarly, MapCoder~\cite{2024_ACL_MapCoder} adapts the human programming cycle to define four key roles for task completion. Another approach, Self-Organized~\cite{2024_arXiv_SelfOrganized}, organizes tasks through a hierarchical parent-child node structure, promoting iterative progress and efficient collaboration.

\subsection{Prompt Engineering}
Prompt engineering plays a crucial role in optimizing code generation tasks by effectively guiding model outputs, ensuring both, consistency and efficiency in the process. Inspired by the CoT (Chain of Thought)~\cite{2022_arXiv_Chain} method, there are usually three main stages in code generation tasks to gradually solve the problem while maintaining clarity and structured reasoning: Planning~\cite{2023_arXiv_MultiAgent,2024_arXiv_Experimenting,2024_arXiv_SOEN101}, Coding~\cite{2024_arXiv_CodePori,2024_arXiv_CodeS,2024_arXiv_MAGISLM}, and Debugging~\cite{2023_arXiv_CAMEL,2024_arXiv_AgentFLSL,2024_arXiv_AIpowered}, AgentCoder~\cite{2023_arXiv_AgentCoder} directs the agent to produce pseudocode following the phases of problem comprehension and algorithm selection. LLM4CBI~\cite{2023_arXiv_Isolating} utilizes a stored component that tracks relevant prompts and selects the most effective ones to guide LLMs in generating variations.

\section{Cogito}

\subsection{Agent Roles}
Building on the "Chain of Thought" (CoT)~\cite{2022_arXiv_Chain} process, we assign three distinct roles within the team: Planner, Coder, and Debugger. The Planner’s role is to outline a clear, step-by-step strategy for solving the problem, considering key aspects such as edge cases and performance issues. This guidance helps the Coder translate the plan into functional code, ensuring that all critical scenarios are addressed during implementation.
After the Coder finishes coding, the solution is tested against a set of sample inputs and expected outputs~\cite{2024_ACL_MapCoder}. If the code passes the tests, it is considered finalized. However, if it fails, the Debugger steps in, analyzing the traceback feedback to identify and correct errors. This collaborative process ensures that the final code is both robust and efficient.

\subsection{Super-Role}
\begin{figure}[t]
    \centering
    \includegraphics[width=1\linewidth]{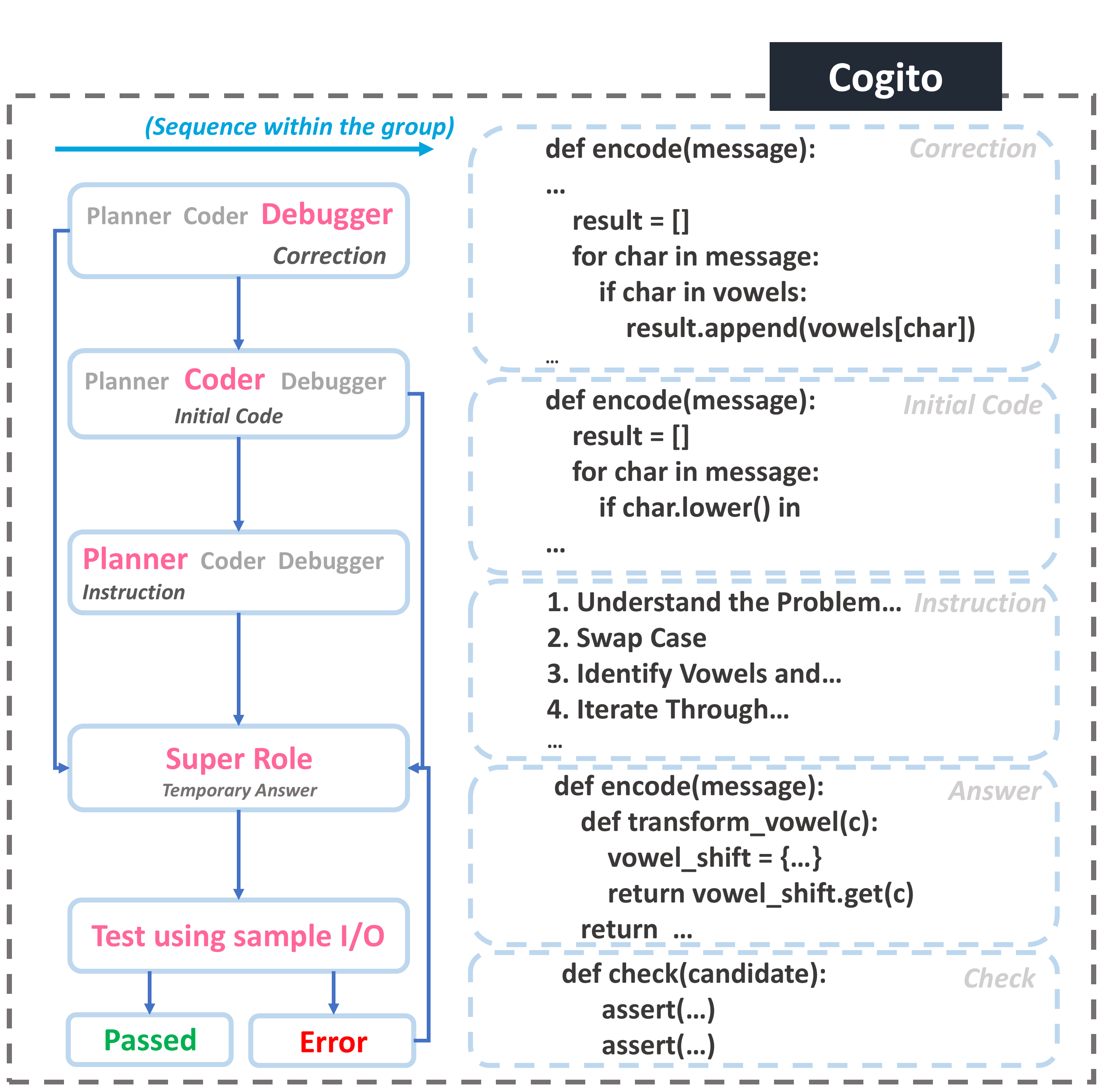}
    \caption{The abbreviated explanation of the process and sample outputs for each step.}
    \label{fig:method_case}
\end{figure}

In this experimental setup, we introduce a shared member known as the \textbf{Super-Role}, who is assigned to each of the three groups sequentially. This member rotates through the roles of Debugger, Coder, and Planner within each group, contributing to a dynamic and collaborative environment. Importantly, the public role retains the memory of all its prior experiences, which plays a crucial role in informing and guiding the execution of its current responsibilities. This memory not only enhances the efficiency of the member’s actions within the same group but also acts as a communication bridge across different groups, facilitating the transfer of knowledge and strategies.
Upon completion of the three distinct tasks, the public member now equipped with accumulated expertise will be entrusted with the task of solving the problem independently. To ensure robustness in the final solution, the member is provided with up to five opportunities for error correction, allowing iterative refinement of the outcome. The final answer, enriched by the cumulative knowledge gained through this process, will be generated and presented by the \textbf{Super-Role}, reflecting its comprehensive learning journey across multiple roles and tasks. A complete example of the response process is shown in Figure~\ref{fig:method_case}.

\subsection{The Hippocampus-like Memory Module}
We drew inspiration from the structural divisions of the human hippocampus~\cite{BURGESS2002625,BERRON2017466,Kesner2013APA} to design our storage module, with each region serving distinct functions. The \textbf{Dentate Gyrus (DG)} is primarily responsible for the formation of new memories and processes a large volume of information. The hippocampus contains different regions of the Cornu Ammonis (CA), each responsible for distinct functions.

\noindent\textbf{CA1 Region.} In this model, we input tasks and corresponding responses into specific regions of the hippocampus, each playing a distinct role in how memories are formed, retained, and recalled over time. The \textbf{CA1} region, pivotal for the storage and retrieval of long-term memories, serves as the repository for initial responses generated during problem-solving. Once formulated, these responses are stored for long-term access and ready for future retrieval when related tasks arise, much like how we retain foundational knowledge and lessons learned over time.

\noindent\textbf{CA2 Region.} While research on the \textbf{CA2} region remains sparse, it is known to play a role in social and emotional memory. This led us to designate it as the "Personalization Module," where users can input their prior code. This personalized memory helps the system understand the user's unique naming conventions and coding practices, guiding the LLM to generate code that aligns with the individual’s style. Though optional, this module ensures that the code generated is tailored to the user’s preferences, fostering a more intuitive and personalized experience.

\noindent\textbf{CA3 Region.} The \textbf{CA3} region, which facilitates quick recall and rapid learning, stores different versions of the code along with the associated error tracebacks. This allows for fast retrieval of past mistakes and corrections, helping avoid errors in future problem-solving processes. This mirrors the brain's ability to learn from past experiences, making future decision-making faster and more efficient.

\noindent\textbf{CA4 Region.} Finally, the \textbf{CA4} region serves as a bridge between the \textbf{DG} and \textbf{CA3}, storing only the final, correct result or the last modified version. This ensures that successful outcomes are quickly accessible for similar tasks, enabling efficient problem-solving and minimizing the time spent on recurring issues. For the Design aspect of the generated answer, we utilize a tokenizer to reduce the size of the stored content, optimizing memory space while retaining essential details for future reference.

\subsection{Agent Collaboration Settings}
To mitigate the potential negative impact of answers generated by different roles during the learning phase, we begin by assigning initial weights to each role, specifically 0.4, 0.4, and 0.3, respectively. Subsequently, the answers generated by each role are evaluated, and an importance score is assigned based on the quality of the results. The final score is derived by multiplying the importance score by the initial weight. This approach ensures that poor answers receive lower scores, while high-quality answers are rewarded with higher scores. As a result, this mechanism not only reduces the influence of incorrect answers but also incentivizes the \textbf{Super-Role} to prioritize and utilize better solutions. Moreover, to further enhance variability and avoid over-committing resources to refine incorrect answers, we reintroduce two randomly selected roles, excluding the one currently used by the public role, in each group. This strategy helps prevent excessive resource allocation to erroneous answers and encourages the generation of more reliable solutions. In each group, the debugger role modifies or improves the code based on the obtained execution results. This process will only be performed once. During the expert phase, the number of debug attempts for experts is uniformly set to 5.
We summarize our agent traversal in Algorithm~\ref{code:Cogito}.


\begin{algorithm}[h]
  \caption{Cogito}
  \label{code:Cogito}
  \begin{algorithmic}[1]
    \State \textbf{Common Agent:} \( \mathit{A}_{\mathit{c}} \), \textbf{Plan Agent:} \( \mathit{A}_{\mathit{p}} \)\\
    \textbf{Implement Agent:} \( \mathit{A}_{\mathit{i}} \), \textbf{Debug Agent:} \( \mathit{A}_{\mathit{d}} \)\\
    \textbf{Super Role:} \( \mathit{S}_{\mathit{R}} \)

    \State Plan\_A $\gets$ \( \mathit{A}_{\mathit{p}} \)(Question)
    \State Code\_A $\gets$ \( \mathit{A}_{\mathit{i}} \)(Plan\_A, Question)
    \State \textbf{Own}\_Answer $\gets$ \( \mathit{A}_{\mathit{c}} \)(Code\_A, sample\_io)

    \State Plan\_B $\gets$ \( \mathit{A}_{\mathit{p}} \)(Question)
    \State \textbf{Own}\_Code $\gets$ \( \mathit{A}_{\mathit{i}} \)(Plan\_B, Question)
    \State Answer\_B $\gets$ \( \mathit{A}_{\mathit{d}} \)(Code\_B, sample\_io)

    \State \textbf{Own}\_Plan $\gets$ \( \mathit{A}_{\mathit{p}} \)(Question)
    \State Code\_C $\gets$ \( \mathit{A}_{\mathit{i}} \)(Plan\_C, Question)
    \State Answer\_C $\gets$ \( \mathit{A}_{\mathit{d}} \)(Code\_C, sample\_io)

    \State tem\_code $\gets$ \( \mathit{S}_{\mathit{R}} \)(Question, \textbf{Own}\_Answer, \textbf{Own}\_Code, \textbf{Own}\_Plan)

    \If {test(tem\_code, sample\_io)}
        \State \textbf{return} tem\_code
    \Else
        \For{$i = 1$ \textbf{to} 5}
            \State code $\gets$ \( \mathit{S}_{\mathit{R}} \)(Question, \textbf{Own}\_Answer, \textbf{Own}\_Code, \textbf{Own}\_Plan)
            \If{test(code, sample\_io)}
                \State \textbf{return} code
            \EndIf
            \State tem\_code $\gets$ code
        \EndFor
        \State \textbf{return} tem\_code
    \EndIf
  \end{algorithmic}
\end{algorithm}


\section{EXPERIMENTS}

\subsection{Experimental Settings}

\textbf{Datasets.} We adopt 8 widely-used benchmark datasets for testing, with 5 datasets containing only simple programming problems (e.g., HumanEval~\cite{2021_arXiv_Evaluating}, HumanEval-ET~\cite{2023_arXiv_CodeScore}, EvalPlus~\cite{2023_is_yourcode}, MBPP~\cite{2021_arXiv_Program}, MBPP-ET~\cite{2023_arXiv_CodeScore}), and others that contain complex programming problems (e.g., Automated Programming Progress Standard (APPS), xCodeEval~\cite{2023_arXiv_XCodeEval}, and CodeContest).
More detailed information is provided in Appendix~\ref{appendix:datasets}.

\noindent\textbf{Evaluation Metric.} For the dataset used in the experiment, we uniformly apply the widely-used unbiased version of \textbf{Pass @k} as evaluation metric~\cite{2021_arXiv_Evaluating,2023_arXiv_Selfcollaboration}.
Note that the unbiased version of \textbf{Pass @k} is a metric used to evaluate recommendation systems by correcting for potential biases in the recommendation process. The formula is given by:

\begin{equation}
\text{Pass@k} = \mathbb{E}_{\text{Problems}} \left[ 1 - \frac{\binom{n-c}{k}}{\binom{n}{k}} \right],
\end{equation}

where \( n \) is the total number of items, \( c \) is the number of relevant items, and \( k \) is the size of the top-k recommendations. 
More detailed information is provided in Appendix~\ref{appendix:evaluation-metric}.


\begin{table*}[t]
\centering
\renewcommand{\arraystretch}{1.2}
\scalebox{0.88}{
    \begin{tabular}{l|l|ccccc|ccc}
    \toprule[2pt]
    \multirow{3}{*}{\textbf{LLM}}     
        & \multicolumn{1}{c}{\multirow{3}{*}{\textbf{Approach}}} 
        & \multicolumn{5}{|c}{\textbf{Simple Problems}}                                                                                                                                         & \multicolumn{3}{|c}{\textbf{Contest-Level Problems}}                              
        \\ \cline{3-10} 
                &                           
                    & \multicolumn{1}{c}{\textbf{HumanEval}} 
                    & \multicolumn{1}{c}{\begin{tabular}[c]{@{}c@{}}\textbf{HumanEval} \\ \textbf{ET}\end{tabular}} 
                    & \multicolumn{1}{c}{\textbf{EvalPlus}}
                    & \multicolumn{1}{c}{\textbf{MBPP}} 
                    & \begin{tabular}[c]{@{}c@{}}\textbf{MBPP} \\ \textbf{ET}\end{tabular} 
                    & \multicolumn{1}{c}{\textbf{APPS}} 
                    & \multicolumn{1}{c}{\textbf{xCodeEval}} & \textbf{CodeContest} 
                \\ \midrule[1pt]
    \multicolumn{1}{c|}{\multirow{9}{*}{\rotatebox{90}{\textbf{ChatGPT-3.5}}}} 
                    & Direct \dag               
                        & \multicolumn{1}{c}{48.1}      
                        & \multicolumn{1}{c}{37.2}                                                    
                        & \multicolumn{1}{c}{66.5}     
                        & \multicolumn{1}{c}{49.8} 
                        & 37.7                                               
                        & \multicolumn{1}{c}{8.0}  
                        & \multicolumn{1}{c}{17.9}      
                        & 5.5         
                    \\ 
                     & CoT \dag                  
                        & \multicolumn{1}{c}{68.9}      
                        & \multicolumn{1}{c}{55.5}                                                    
                        & \multicolumn{1}{c}{65.2}     
                        & \multicolumn{1}{c}{54.5} 
                        & 39.6                                               
                        & \multicolumn{1}{c}{7.3}  
                        & \multicolumn{1}{c}{23.6}      
                        & 6.1         
                    \\ 
                    & Self-Planning \dag        
                        & \multicolumn{1}{c}{60.3}      
                        & \multicolumn{1}{c}{46.2}                                                    
                        & \multicolumn{1}{c}{-}        
                        & \multicolumn{1}{c}{55.7} 
                        & 41.9                                               
                        & \multicolumn{1}{c}{9.3}  
                        & \multicolumn{1}{c}{18.9}      
                        & 6.1         
                    \\ 
                    & Analogical \dag           
                        & \multicolumn{1}{c}{63.4}      
                        & \multicolumn{1}{c}{50.6}                                                    
                        & \multicolumn{1}{c}{59.1}     
                        & \multicolumn{1}{c}{70.5} 
                        & 46.1                                               
                        & \multicolumn{1}{c}{6.7}  
                        & \multicolumn{1}{c}{15.1}      
                        & 7.3         
                    \\ 
                    & Reflexion \dag            
                        & \multicolumn{1}{c}{67.1}      
                        & \multicolumn{1}{c}{49.4}                                                    
                        & \multicolumn{1}{c}{62.2}     
                        & \multicolumn{1}{c}{73.0} 
                        & 47.4                                               
                        & \multicolumn{1}{c}{-}    
                        & \multicolumn{1}{c}{-}         
                        & -           
                    \\ 
                    & Self-Collaboration \dag   
                        & \multicolumn{1}{c}{74.4}      
                        & \multicolumn{1}{c}{56.1}                                                    
                        & \multicolumn{1}{c}{-}        
                        & \multicolumn{1}{c}{68.2} 
                        & 49.5                                               
                        & \multicolumn{1}{c}{-}    
                        & \multicolumn{1}{c}{-}         
                        & -           
                    \\ 
                    & MapCoder \dag             
                        & \multicolumn{1}{c}{\textcolor{mycolor_3}{\underline{80.5}}}      
                        & \multicolumn{1}{c}{\textcolor{mycolor_3}{\underline{70.1}}}                                                    
                        & \multicolumn{1}{c}{\textcolor{mycolor_3}{\underline{71.3}}} 
                        & \multicolumn{1}{c}{\textcolor{mycolor_3}{\underline{78.3}}} 
                        & \textcolor{mycolor_3}{\underline{54.4}}                                               
                        & \multicolumn{1}{c}{\textcolor{mycolor_3}{\underline{11.3}}} 
                        & \multicolumn{1}{c}{\textcolor{mycolor_3}{\underline{27.4}}}      
                        & \textcolor{mycolor_3}{\underline{12.7}}        
                    \\ 
                    \cline{2-10} 
                    \multirow{2}{*}{} & \multirow{2}{*}{\textbf{\textcolor{mycolor_2}{Cogito (Ours)}}}  
                        & \multicolumn{1}{c}{\multirow{2}{*}{\textbf{\makecell{\textcolor{mycolor_2}{90.24} \\ \textcolor{red}{$\uparrow$ 12.1\%}}}}}
                        & \multicolumn{1}{c}{\multirow{2}{*}{\textbf{\makecell{\textcolor{mycolor_2}{81.71} \\ \textcolor{red}{$\uparrow$ 16.6\%}}}}} 
                        & \multicolumn{1}{c}{\multirow{2}{*}{\textbf{\makecell{\textcolor{mycolor_2}{82.32} \\ \textcolor{red}{$\uparrow$ 15.5\%}}}}}
                        & \multicolumn{1}{c}{\multirow{2}{*}{\textbf{\makecell{\textcolor{mycolor_2}{85.14} \\ \textcolor{red}{$\uparrow$ 8.7\%}}}}}
                        & \multicolumn{1}{c}{\multirow{2}{*}{\textbf{\makecell{\textcolor{mycolor_2}{59.70} \\ \textcolor{red}{$\uparrow$ 9.7\%}}}}}
                        & \multicolumn{1}{|c}{\multirow{2}{*}{\textbf{\makecell{\textcolor{mycolor_2}{18.00} \\ \textcolor{red}{$\uparrow$ 59.29\%}}}}}
                        & \multicolumn{1}{c}{\multirow{2}{*}{\textbf{\makecell{\textcolor{mycolor_2}{30.19} \\ \textcolor{red}{$\uparrow$ 10.2\%}}}}}
                        & \multicolumn{1}{c}{\multirow{2}{*}{\textbf{\makecell{\textcolor{mycolor_2}{13.33} \\ \textcolor{red}{$\uparrow$ 5.0\%}}}}}
                    \\  
                        & \multicolumn{1}{c}{}
                        & \multicolumn{1}{|c}{}
                        & \multicolumn{1}{c}{}
                        & \multicolumn{1}{c}{}
                        & \multicolumn{1}{c}{}
                        & \multicolumn{1}{c}{}
                        & \multicolumn{1}{|c}{}
                        & \multicolumn{1}{c}{}
                        & \multicolumn{1}{c}{}
                    \\ \midrule[1pt]
    \multicolumn{1}{c|}{\multirow{8}{*}{\rotatebox{90}{\textbf{GPT-4}}}}   
                    & Direct \dag               
                        & \multicolumn{1}{c}{80.1}      
                        & \multicolumn{1}{c}{73.8}                                                    
                        & \multicolumn{1}{c}{81.7}     
                        & \multicolumn{1}{c}{81.1} 
                        & 54.7                                               
                        & \multicolumn{1}{c}{12.7} 
                        & \multicolumn{1}{c}{32.1}      
                        & 12.1        
                    \\ 
                    & CoT \dag                  
                        & \multicolumn{1}{c}{89.0}      
                        & \multicolumn{1}{c}{61.6}                                                    
                        & \multicolumn{1}{c}{-}        
                        & \multicolumn{1}{c}{82.4} 
                        & 56.2                                               
                        & \multicolumn{1}{c}{11.3} 
                        & \multicolumn{1}{c}{36.8}      
                        & 5.5         
                    \\ 
                    & Self-Planning \dag        
                        & \multicolumn{1}{c}{85.4}      
                        & \multicolumn{1}{c}{62.2}                                                    
                        & \multicolumn{1}{c}{-}        
                        & \multicolumn{1}{c}{75.8} 
                        & 50.4                                               
                        & \multicolumn{1}{c}{14.7} 
                        & \multicolumn{1}{c}{34.0}      
                        & 10.9        
                    \\ 
                    & Analogical \dag           
                        & \multicolumn{1}{c}{66.5}      
                        & \multicolumn{1}{c}{48.8}                                                    
                        & \multicolumn{1}{c}{62.2}     
                        & \multicolumn{1}{c}{58.4} 
                        & 40.3                                               
                        & \multicolumn{1}{c}{12.0}
                        & \multicolumn{1}{c}{26.4}      
                        & 10.9        
                    \\ 
                    & Reflexion \dag            
                        & \multicolumn{1}{c}{91.0}      
                        & \multicolumn{1}{c}{78.7}                                                    
                        & \multicolumn{1}{c}{81.7}     
                        & \multicolumn{1}{c}{78.3} 
                        & 51.9                                               
                        & \multicolumn{1}{c}{-}    
                        & \multicolumn{1}{c}{-}         
                        & -           
                    \\ 
                    & MapCoder \dag             
                        & \multicolumn{1}{c}{\textcolor{mycolor_3}{\underline{93.9}}}      
                        & \multicolumn{1}{c}{\textcolor{mycolor_3}{\underline{82.9}}}                                                    
                        & \multicolumn{1}{c}{\textcolor{mycolor_3}{\underline{83.5}}}     
                        & \multicolumn{1}{c}{\textcolor{mycolor_3}{\underline{83.1}}} 
                        & \textcolor{mycolor_3}{\underline{57.7}}                                               
                        & \multicolumn{1}{c}{\textcolor{mycolor_3}{\underline{22}}}   
                        & \multicolumn{1}{c}{\textcolor{mycolor_3}{\underline{45.3}}}      
                        & \textcolor{mycolor_3}{\underline{28.5}}        
                    \\ 
                    \cline{2-10} 
                    \multirow{2}{*}{} & \multirow{2}{*}{\textbf{\textcolor{mycolor_2}{Cogito (Ours)}}}  
                        & \multicolumn{1}{c}{\multirow{2}{*}{\textbf{\makecell{\textcolor{mycolor_2}{95.73} \\ \textcolor{red}{$\uparrow$ 1.9\%}}}}}
                        & \multicolumn{1}{c}{\multirow{2}{*}{\textbf{\makecell{\textcolor{mycolor_2}{83.54} \\ \textcolor{red}{$\uparrow$ 0.8\%}}}}} 
                        & \multicolumn{1}{c}{\multirow{2}{*}{\textbf{\makecell{\textcolor{mycolor_2}{85.37} \\ \textcolor{red}{$\uparrow$ 2.2\%}}}}}
                        & \multicolumn{1}{c}{\multirow{2}{*}{\textbf{\makecell{\textcolor{mycolor_2}{88.16} \\ \textcolor{red}{$\uparrow$ 6.1\%}}}}}
                        & \multicolumn{1}{c}{\multirow{2}{*}{\textbf{\makecell{\textcolor{mycolor_2}{66.25} \\ \textcolor{red}{$\uparrow$ 14.8\%}}}}}
                        & \multicolumn{1}{|c}{\multirow{2}{*}{\textbf{\makecell{\textcolor{mycolor_2}{27.30} \\ \textcolor{red}{$\uparrow$ 24.1\%}}}}}
                        & \multicolumn{1}{c}{\multirow{2}{*}{\textbf{\makecell{\textcolor{mycolor_2}{47.20} \\ \textcolor{red}{$\uparrow$ 4.2\%}}}}}
                        & \multicolumn{1}{c}{\multirow{2}{*}{\textbf{\makecell{\textcolor{mycolor_2}{29.70} \\ \textcolor{red}{$\uparrow$ 4.2\%}}}}}
                    \\  
                        & \multicolumn{1}{c}{}
                        & \multicolumn{1}{|c}{}
                        & \multicolumn{1}{c}{}
                        & \multicolumn{1}{c}{}
                        & \multicolumn{1}{c}{}
                        & \multicolumn{1}{c}{}
                        & \multicolumn{1}{|c}{}
                        & \multicolumn{1}{c}{}
                        & \multicolumn{1}{c}{}
                    \\ \bottomrule[2pt]
    \end{tabular}
}
\caption{Overall performance comparison across various datasets, categorized into \textit{Simple Problems} and \textit{Contest-Level Problems}.  The best performance is highlighted in \textbf{\textcolor{mycolor_2}{blue}}, and the second-best performance is \textcolor{mycolor_3}{\underline{underlined}}. The improvement of \textbf{Cogito} over the previous state-of-the-art (Mapcoder) is highlighted in \textbf{\textcolor{red}{red}}. \dag: Results are publicly disclosed in the paper of MapCoder.}
\label{tab_Overall-Performance}
\end{table*}

\noindent\textbf{Baselines.}
We conduct a comprehensive comparison with several representative methods: Direct, Chain-of-Thought (CoT), Self-Planning, Analogical Reasoning, and MapCoder.
More detailed information is provided in Appendix~\ref{appendix:baselines}.

\subsection{Overall Performance}
In this section, we conduct a comprehensive evaluation of our proposed process, and all the results are systematically presented in Table~\ref{tab_Overall-Performance}. From the table, it is evident that \textbf{Cogito} outperforms all the other models, achieving the highest scores across all datasets. Notably, the application of GPT-4 significantly enhances the overall performance, yielding the best results observed in our experiments. These results underscore the effectiveness of \textbf{Cogito} and highlight the substantial improvements brought by the integration of GPT-4 into the process, demonstrating its potential for high-level performance across diverse data scenarios. The results further validate the effectiveness of our growth-based learning concept, demonstrating that enabling the agent to evolve reversely can enhance its problem-solving capabilities.

\subsubsection{Performance on Simple Code Generation Tasks}
Table~\ref{tab_Overall-Performance} presents the performance results of various baseline methods, along with the corresponding percentage improvements achieved by our method when compared to the direct prompting approach. Our proposed method, Cogito, outperforms all other approaches in each of the experiments conducted. Specifically, when compared to the current state-of-the-art model, MapCoder, Cogito demonstrates significant improvements in Pass@1 accuracy across multiple datasets: 12.1\%, 16.6\%, 15.5\%, 8.7\%, and 9.7\% on HumanEval, HumanEval-ET, EvalPlus, MBPP, and MBPP-ET, respectively, when utilizing GPT-3.5-turbo. Notably, in comparison to the direct prompting method, Cogito achieves an impressive improvement of up to 119.65\%. Furthermore, by leveraging the capabilities of GPT-4, the performance across all datasets saw a substantial enhancement, culminating in the highest values recorded in our experiments.


\begin{table*}[]
\centering
\renewcommand{\arraystretch}{1.23}
\scalebox{0.74}{
    \begin{tabular}{c|l|cccc|cc|cc}
    \toprule[2pt]
    \multicolumn{1}{c|}{\multirow{2}{*}{\textbf{LLM}}}         
        & \multicolumn{1}{c}{\multirow{2}{*}{\textbf{Dataset}}} 
        & \multicolumn{4}{|c}{\textbf{Average for Cogito}}                                                                                                 
        & \multicolumn{2}{|c}{\textbf{\makecell{Average for\\MapCoder}}}  
        & \multicolumn{1}{|l}{\textbf{\multirow{2}{*}{\makecell{API Calls\\Reduction}}}}
        & \multicolumn{1}{l}{\textbf{\multirow{2}{*}{\makecell{Token\\Reduction(k)}}}} 
        \\ 
        \cline{3-8} 
    \multicolumn{1}{c|}{}                             
        & \multicolumn{1}{c}{}                         
        & \multicolumn{1}{|c}{\textbf{\makecell{API Calls}}} 
        & \multicolumn{1}{c}{\textbf{\makecell{Prompt Tokens (k)}}} 
        & \multicolumn{1}{c}{\textbf{\makecell{Completion Tokens (k)}}} 
        & \textbf{\makecell{Total Tokens (k)}} 
        & \multicolumn{1}{c}{\textbf{\makecell{API Calls}}} 
        & \multicolumn{1}{c}{\textbf{Tokens (k)}} 
        & \multicolumn{1}{|c}{}
        & \multicolumn{1}{c}{}                    
        \\ 
        \midrule[1pt]
    \multicolumn{1}{c|}{\multirow{5}{*}{\rotatebox{90}{\textbf{ChatGPT-3.5}}}} 
        & HumanEval                                     
        & \multicolumn{1}{c}{10}        
        & \multicolumn{1}{c}{4.95}              
        & \multicolumn{1}{c}{1.31}                  
        & 6.26             
        & \multicolumn{1}{c}{17}         
        & \multicolumn{1}{c}{10.41}
        & \multicolumn{1}{|c}{7}
        & \multicolumn{1}{c}{4.15}                    
        \\ 
    \multicolumn{1}{c|}{}                             
        & MBPP                                          
        & \multicolumn{1}{c}{9}         
        & \multicolumn{1}{c}{3.28}              
        & \multicolumn{1}{c}{1.32}                  
        & 4.60             
        & \multicolumn{1}{c}{12}         
        & \multicolumn{1}{c}{4.84}
        & \multicolumn{1}{|c}{3}
        & \multicolumn{1}{c}{0.24}                    
        \\ 
    \multicolumn{1}{c|}{}                             
        & APPS                                          
        & \multicolumn{1}{c}{14}        
        & \multicolumn{1}{c}{14.74}             
        & \multicolumn{1}{c}{2.04}                  
        & 16.78            
        & \multicolumn{1}{c}{21}         
        & \multicolumn{1}{c}{26.57}
        & \multicolumn{1}{|c}{7}
        & \multicolumn{1}{c}{2.04}                    
        \\ 
    \multicolumn{1}{c|}{}                            
        & xCodeEval                                     
        & \multicolumn{1}{c}{16}        
        & \multicolumn{1}{c}{16.21}             
        & \multicolumn{1}{c}{2.69}                  
        & 18.90            
        & \multicolumn{1}{c}{19}         
        & \multicolumn{1}{c}{24.10}
        & \multicolumn{1}{|c}{3}
        & \multicolumn{1}{c}{7.89}                    
        \\ 
    \multicolumn{1}{c|}{}                             
        & CodeContest                                   
        & \multicolumn{1}{c}{15}        
        & \multicolumn{1}{c}{24.54}             
        & \multicolumn{1}{c}{4.02}                  
        & 28.56            
        & \multicolumn{1}{c}{23}         
        & \multicolumn{1}{c}{34.95}
        & \multicolumn{1}{|c}{8}
        & \multicolumn{1}{c}{6.39}                    
        \\ 
        \midrule[0.5pt]
    \multicolumn{1}{c|}{\multirow{5}{*}{\rotatebox{90}{\textbf{GPT-4}}}}       
        & HumanEval                                     
        & \multicolumn{1}{c}{10}          
        & \multicolumn{1}{c}{5.40}                  
        & \multicolumn{1}{c}{1.70}                      
        & 7.10                  
        & \multicolumn{1}{c}{15}         
        & \multicolumn{1}{c}{12.75}
        & \multicolumn{1}{|c}{5}
        & \multicolumn{1}{c}{5.65}                    
        \\ 
    \multicolumn{1}{c|}{}                             
        & MBPP                                          
        & \multicolumn{1}{c}{10}          
        & \multicolumn{1}{c}{3.99}                  
        & \multicolumn{1}{c}{1.58}                      
        & 4.80                  
        & \multicolumn{1}{c}{8}         
        & \multicolumn{1}{c}{4.96}
        & \multicolumn{1}{|c}{-2}
        & \multicolumn{1}{c}{0.16}                    
        \\ 
    \multicolumn{1}{c|}{}                             
        & APPS                                          
        & \multicolumn{1}{c}{13}          
        & \multicolumn{1}{c}{18.20}                  
        & \multicolumn{1}{c}{3.76}                      
        & 21.96                  
        & \multicolumn{1}{c}{19}         
        & \multicolumn{1}{c}{31.8}
        & \multicolumn{1}{|c}{6}
        & \multicolumn{1}{c}{9.84}                    
        \\ 
    \multicolumn{1}{c|}{}                             
        & xCodeEval                                     
        & \multicolumn{1}{c}{14}          
        & \multicolumn{1}{c}{14.01}                  
        & \multicolumn{1}{c}{3.91}                      
        & 17.93                  
        & \multicolumn{1}{c}{14}         
        & \multicolumn{1}{c}{23.45}
        & \multicolumn{1}{|c}{0}
        & \multicolumn{1}{c}{5.52}                    
        \\ 
    \multicolumn{1}{c|}{}                             
        & CodeContest                                   
        & \multicolumn{1}{c}{15}          
        & \multicolumn{1}{c}{26.72}                  
        & \multicolumn{1}{c}{5.63}                      
        & 32.35                  
        & \multicolumn{1}{c}{19}         
        & \multicolumn{1}{c}{38.7}
        & \multicolumn{1}{|c}{4}
        & \multicolumn{1}{c}{6.35}                    
        \\ 
        \hline
    \multicolumn{2}{c}{\textbf{Average}}                                                                      
        & \multicolumn{1}{|c}{\textbf{\textcolor{mycolor_2}{13}}}          
        & \multicolumn{1}{c}{13.20}                  
        & \multicolumn{1}{c}{2.80}                      
        & \textbf{\textcolor{mycolor_2}{16.00}}                  
        & \multicolumn{1}{c}{\textcolor{mycolor_3}{\underline{16.7}}}         
        & \multicolumn{1}{c}{\textcolor{mycolor_3}{\underline{21.25}}} 
        & \multicolumn{1}{|c}{\textbf{\textcolor{red}{4.1}}}
        & \multicolumn{1}{c}{\textbf{\textcolor{red}{4.82}}}                    
        \\ 
    \bottomrule[2pt]
    \end{tabular}
}
\caption{The number of API calls and token consumption for different tasks, compared to the usage reduction with MapCoder.}
\label{tab:API-Token}
\end{table*}

\subsubsection{Performance on Complex Code Generation Tasks}
Contest-level problems feature more comprehensive problem descriptions and a greater number of test cases, with no limitation on the generation of a single function to address the task. Cogito has shown notable advancements compared to MapCoder across most datasets, including APPS, xCodeEval, and CodeContests. Specifically, for GPT-3.5-turbo, improvements of 59.29\%, 30.19\%, and 5.0\% are observed, while for GPT-4, the enhancements are 24.1\%, 4.2\%, and 4.2\%, respectively. Adhering to the unified testing validation approach, we require that all responses within these three datasets be implemented as functions that take a string parameter, returning the result as a string via the `return` statement. Despite the advantages of this methodology, its application has led to a decline in performance on certain platforms, notably xCodeEval and CodeContests.

\begin{figure}
    \centering
   \begin{minipage}{0.49\linewidth}
        \centering
        \includegraphics[width=\linewidth]{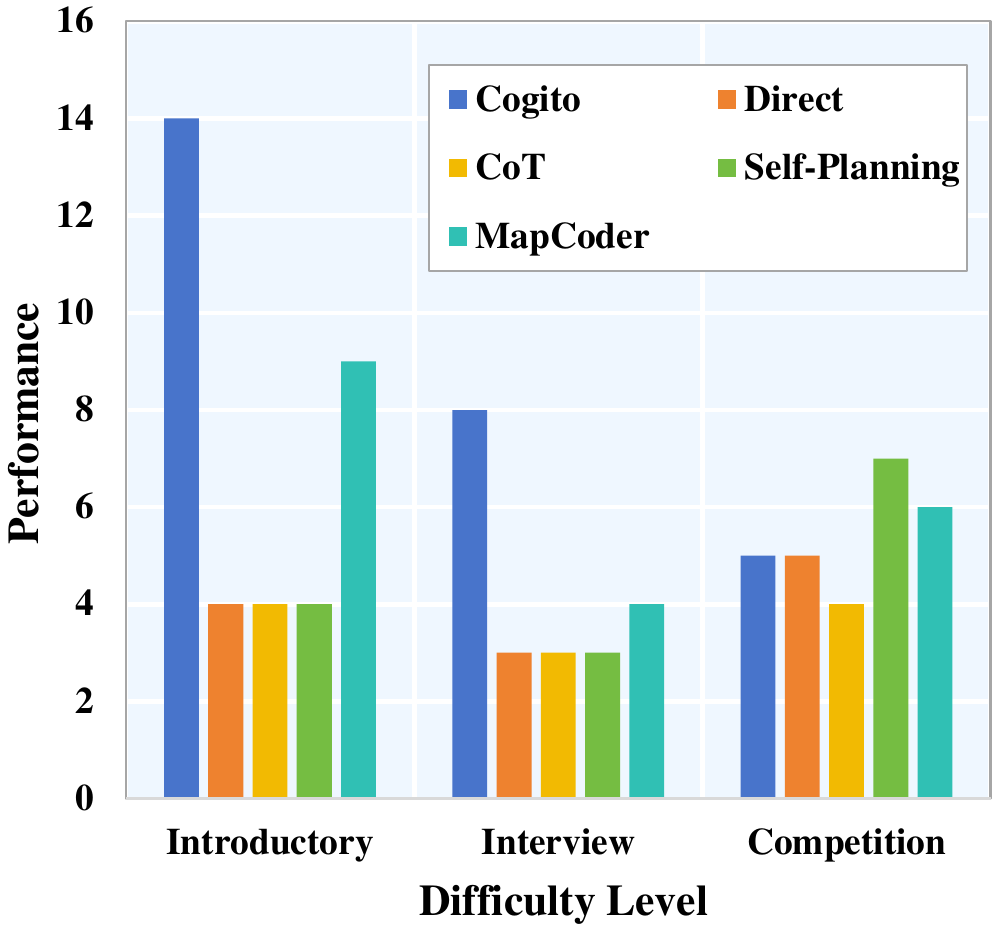}
        \caption*{(a)}
    \end{minipage}%
    \begin{minipage}{0.49\linewidth}
        \centering
        \includegraphics[width=\linewidth]{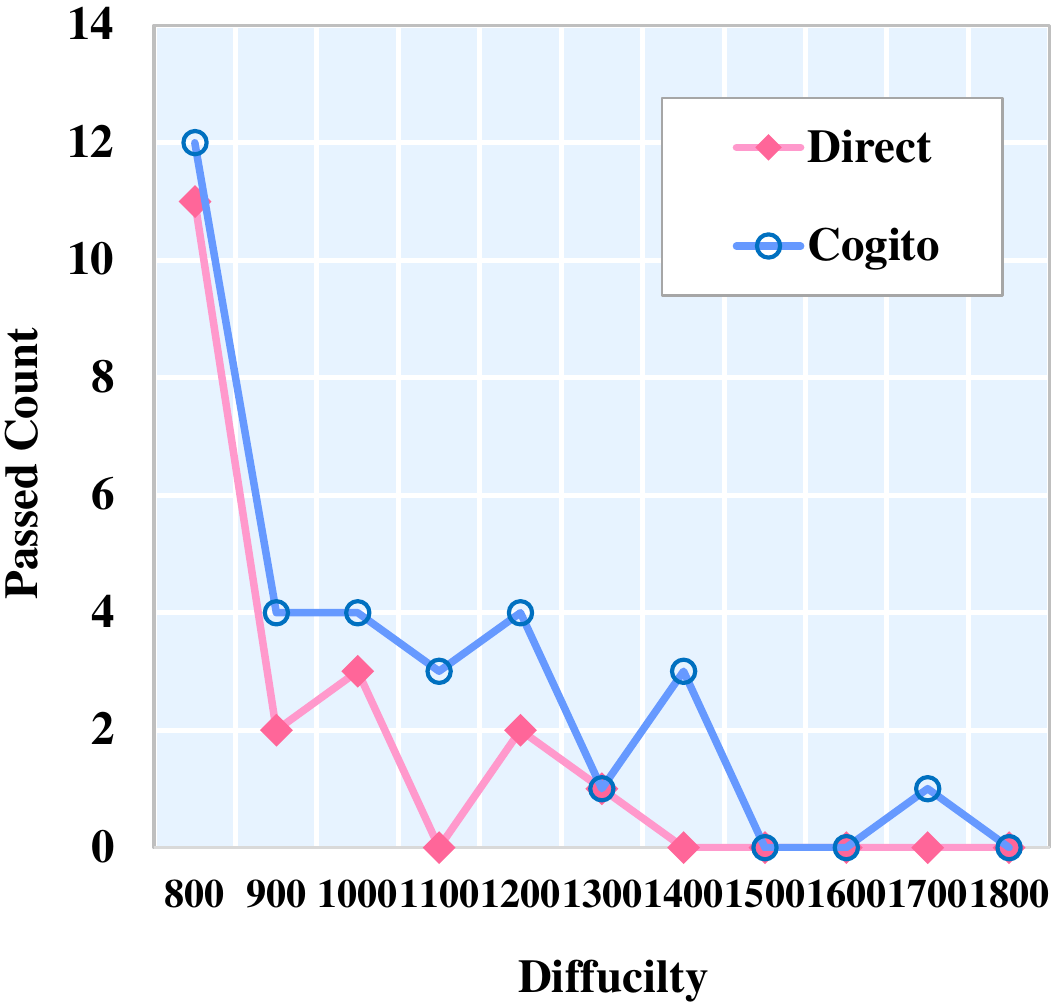}
        \caption*{(b)}
    \end{minipage}
    \caption{The comparison results on representative datasets.}
    \label{fig:difficulty_Performance}
\end{figure}

\begin{figure}[!h]
    \centering
    \includegraphics[width=0.55\linewidth]{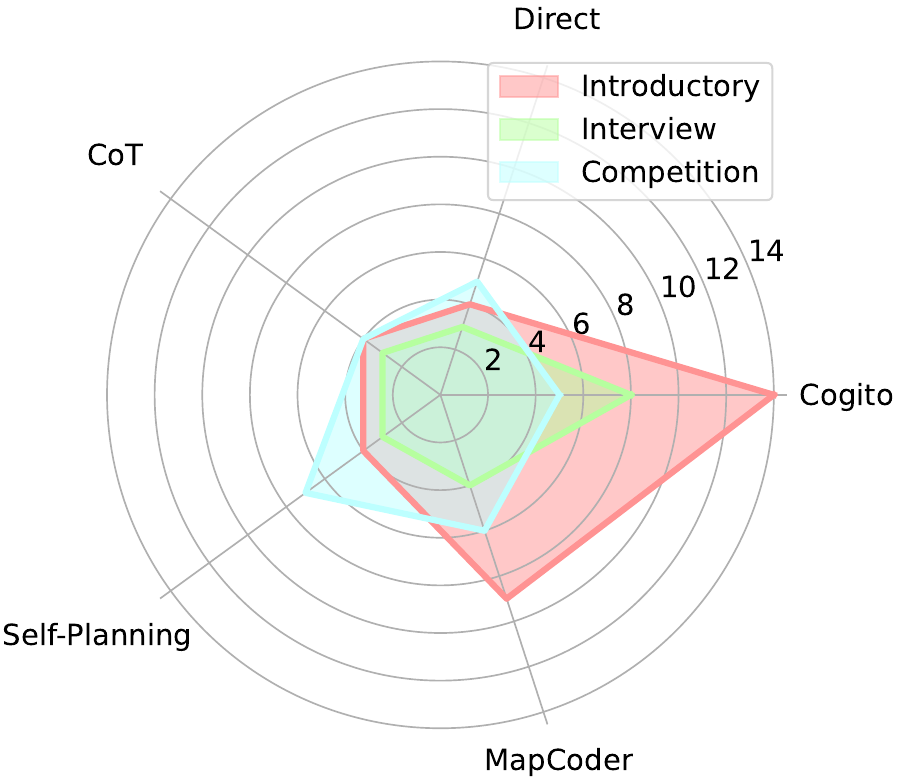}
    \caption{The comparison results with respect to the algorithm and difficulty levels (APPS dataset).}
    \label{fig:Difficulty-Level_Performance}
\end{figure}

\subsubsection{Performance Under Different Difficulty Levels}

\textbf{Difficulty levels.} The APPS dataset consists of problems with three difficulty levels: (i) Introductory, (ii) Interview, and (iii) Competition. Figure~\ref{fig:difficulty_Performance}(a) and Figure~\ref{fig:Difficulty-Level_Performance} show the number of problems solved by different methods at different levels under these three classifications. At the Introductory and Interview levels, Cogito significantly outperforms existing methods, highlighting the effectiveness of our approach for relatively simple and moderately difficult code generation tasks.

\noindent\textbf{Difficulty score.} 
In the xCodeEval dataset, each task is assigned a difficulty score, with the difficulty scores of the answers that successfully pass the tests ranging from 800 to 1800. We compare our approach with the direct method, and our results consistently outperforme the direct method across different difficulty levels (Figure~\ref{fig:difficulty_Performance}(b)).

\subsubsection{Consumption of API and Tokens}

In Table~\ref{tab:API-Token}, we present the number of API calls and tokens (measured in k) utilized by both GPT-3.5-turbo and GPT-4 across different datasets. Specifically, for the HumanEval dataset, we observe a remarkable reduction of up to 66.29\% in token usage and 70\% in API calls. On average, token consumption decreases by 32.81\%, and API calls are reduced by 28.46\%. However, it is important to highlight that, due to GPT-4’s tendency to generate more detailed explanatory content, the overall resource consumption increases compared to GPT-3.5-turbo, possibly due to the model’s structure.

\subsection{Ablation Study}


\begin{table}[t]
\centering
\renewcommand{\arraystretch}{1.3}
\scalebox{0.8}{
    \begin{tabular}{l|cc}
    \toprule[2pt]
    \multicolumn{1}{c|}{\textbf{Model}}          & \textbf{Pass@1} & \textbf{\makecell{Perforcemance\\Drop}} 
        \\ 
        \midrule[1pt]
    \textbf{Cogito w/o Planning Experience}         &   76.22              &         14.02                    
        \\ 
    \textbf{Cogito w/o Implementation Experience} &   78.05              &         12.19                   
        \\ 
    \textbf{Cogito w/o Debugging Experience}           &   79.88              &         10.36                    
        \\ 
    \textbf{Cogito w/o Super-Role}                &   69.33              &         20.91                    
        \\ 
    \textbf{Normal Sequence}                    &   73.17              &         17.07                    
    \\ 
    \bottomrule[2pt]
    \end{tabular}
}
\caption{Ablation study results on HumanEval using GPT-3.5-turbo. The table shows the impact of different components or configurations on performance.}
\label{Ablation_diff_role}
\end{table}

\textbf{Impact of Different Agents.} 
To verify the effectiveness of the proposed approach, we systematically remove the active participation of various key roles involved in the process. The experimental results (Table~\ref{Ablation_diff_role}) indicate that omitting the critical planning phase led to a maximum performance drop of 14.02\%. The absence of hands-on Implementation practice reduces performance by 12.19\%, while the lack of expert Debugging knowledge causes a 10.36\% decline.

\noindent\textbf{Impact of Work Sequence.} 
To rigorously assess the distinctions between our work and previous approaches, particularly in terms of the sequence of experience accumulation, we conduct experiments in which we systematically alter the order in which experiences accumulate. Initially, we employ a sequence where the planner is introduced first, followed by the coder, and finally the debugger. Upon analyzing the results (Table~\ref{Ablation_diff_role}), a performance drop of 17.07\% clearly indicates that the altered sequence contributes to a significant deterioration in the final outcome.

\noindent\textbf{Impact of Super-Role.} 
In the third group, common roles are actively involved in every stage of the process. However, does this justify relying on them for the final answer? Table~\ref{Ablation_diff_role} presents the results. Even with extensive experience and insightful recommendations, planners may still fail to achieve desired outcomes due to a lack of hands-on experience or expertise in addressing practical challenges during implementation. Thus, the \textbf{Super-Role}'s final answer is indispensable.

\noindent\textbf{Impact of Sample I/O.}
In this study, we augment the HumanEval dataset with input-output pairs from MapCoder~\cite{2024_ACL_MapCoder} dataset and five additional test cases from HumanEval-ET dataset. Our results show a modest 0.6\% improvement, indicating a slight positive effect on model performance. These findings suggest that dataset augmentation with diverse test cases can improve model accuracy.

\subsection{Hyper-parameter Analysis}

\textbf{Impact of \textit{t}.}
\begin{table}[t]
\centering
\renewcommand{\arraystretch}{1.3}
\scalebox{0.8}{
    \begin{tabular}{p{0.19\textwidth}|>{\centering\arraybackslash}p{0.12\textwidth}>{\centering\arraybackslash}p{0.12\textwidth}}
    \toprule[2pt]
    \multicolumn{1}{c|}{\multirow{2}{*}{\textbf{Dataset}}} & \multicolumn{2}{c}{\textbf{Debug Times(t)}}                                      
        \\ 
        \cline{2-3} 
                                      & \textbf{3} & \textbf{5}            
        \\ 
        \midrule[1pt]
    \textbf{HumanEval}                
        & 86.59           
        & 90.24                 
        \\ 
    \textbf{HumanEval-ET}             
        & 78.05           
        & 81.71                 
        \\ 
        \bottomrule[2pt]
    \end{tabular}
}
\caption{Analysis of debugging times on representative datasets.}
\label{debug_times}
\end{table}

It involves a single hyperparameter: the number of self-debugging attempts, denoted as \textit{t}. As shown in Table~\ref{debug_times}, increasing the value of $t$ improves the performance. However, this enhancement comes with a trade-off, as it requires more computational time and an increased number of tokens to complete the process. This observation highlights the inherent balance between performance and resource consumption in the proposed method.

\noindent\textbf{Impact of the Number of Iterations for Accumulating Experience.} 
Initially, we set the number of roles to 3, indicating that we require three groups per experience-learning cycle. Increasing the number to six seems intuitive for better experience accumulation. However, as a detailed comparison provided in Table~\ref{member_of_groups}, increasing the number actually leads to decreased performance and higher token consumption. Such results indicate that setting one experience-learning cycle for Cogito is enough and reasonable for improving performance.

\begin{table}[t]
\centering
\renewcommand{\arraystretch}{1.3}
\scalebox{0.8}{
    \begin{tabular}{p{0.19\textwidth}|>{\centering\arraybackslash}p{0.12\textwidth}>{\centering\arraybackslash}p{0.12\textwidth}}
    \toprule[2pt]
    \multicolumn{1}{c|}{\multirow{2}{*}{\textbf{Number of Groups}}} & \multicolumn{2}{c}{\textbf{Results}}                                      
        \\ 
        \cline{2-3} 
                                      & \textbf{Pass@1} & \textbf{Average Token(k)}            
        \\ 
        \midrule[1pt]
    \textbf{Three Groups}                
        & 90.24           
        & 10.41                 
        \\ 
    \textbf{Six Groups}             
        & 80.49           
        & 17.43                
        \\ 
        \bottomrule[2pt]
    \end{tabular}
}
\caption{Analysis of debugging times on representative datasets.}
\label{member_of_groups}
\end{table}

\subsection{Case Study}
\subsubsection{New Random Roles vs. Same Roles}

\begin{figure}[t]
    \centering
    \includegraphics[width=1\linewidth]{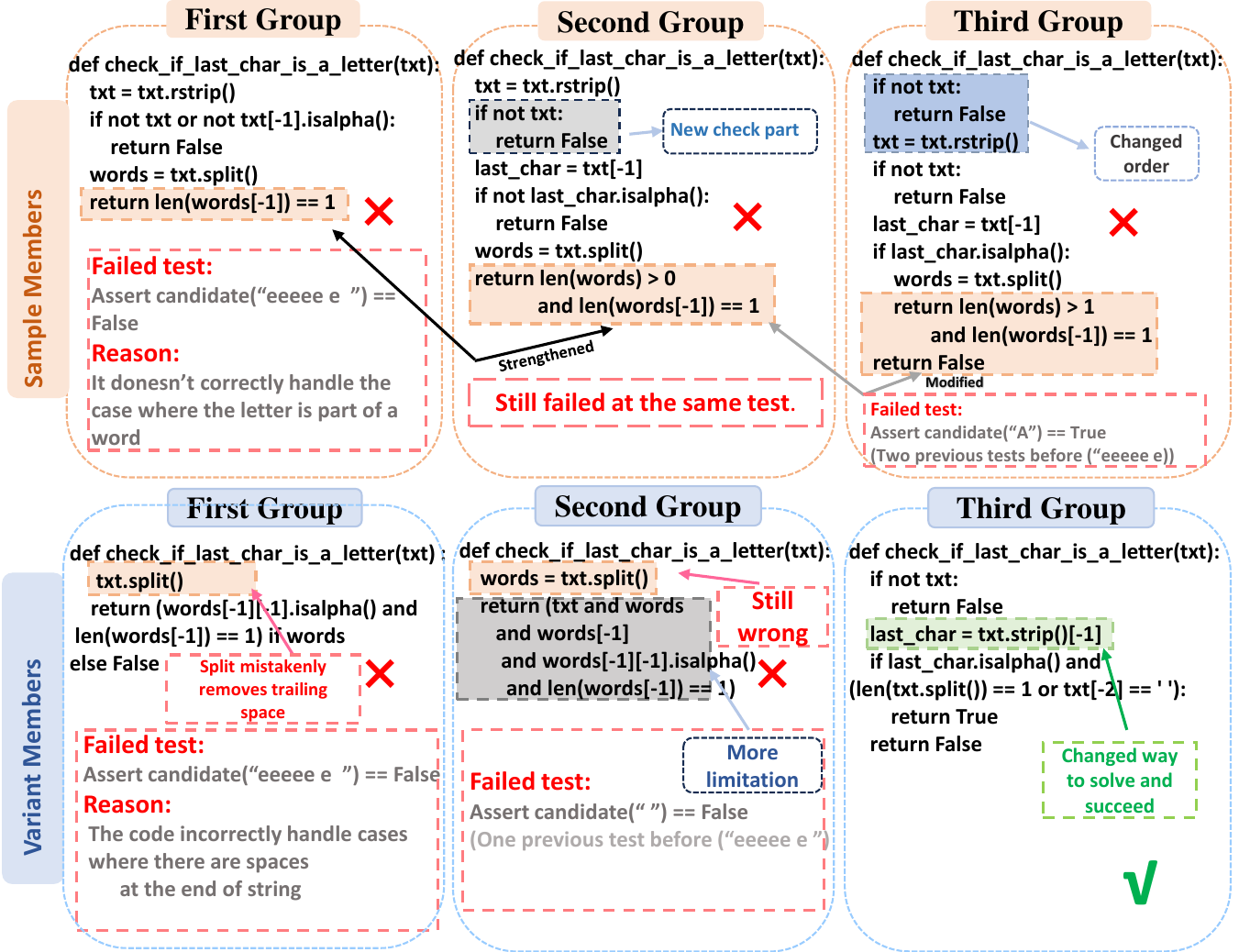}
    \caption{An example of answers from the same group members and different group members on HumanEval.}
    \label{fig:case-study}
\end{figure}

In group discussion sessions, each group will reintroduce two new random members to assume two distinct roles. The question arises: why not allow the same two members to continue participating throughout the role transitions? The rationale behind this decision lies in the potential pitfalls of starting with an incorrect approach. If the direction of code writing is flawed in the initial phase, any subsequent improvements or redesigns will be built upon this foundational error. No matter how many times debugging is performed, the final result will inevitably remain compromised. Therefore, it is essential to ensure that the foundation is correct before moving forward with further development. An example is shown in Figure~\ref{fig:case-study}, demonstrating the necessity of introducing new random roles.

\section{Conclusion}
In this work, we introduce \textbf{Cogito}, a neurobiologically-inspired multi-agent framework for code generation that redefines the traditional workflow of planning, coding, and debugging by adopting a reverse approach. By mimicking the human growth process, \textbf{Cogito} progressively develops its capabilities, transitioning through specialized roles—Debugger, Coder, and Planner—and ultimately evolving into a \textbf{Super-Role} capable of autonomously handling complex code generation tasks. Through extensive evaluations on multiple representative datasets, \textbf{Cogito} demonstrates its ability to achieve state-of-the-art performance with higher efficiency and lower computational cost compared to existing methods. These results highlight the potential of biologically-inspired design principles in advancing intelligent systems. Future work will focus on further optimizing \textbf{Cogito}’s architecture and exploring its applicability to broader software engineering tasks.

\clearpage

\bibliographystyle{named}
\bibliography{main}

\appendix

\onecolumn

\section*{\centering \LARGE \includegraphics[width=0.06\linewidth]{figs/logo_3-2.png} Cogito, ergo sum: A Neurobiologically-Inspired Cognition-Memory-Growth System for Code Generation}
\section*{\centering \LARGE \textmd{Appendix}}

\section{Agent Prompt Details}\label{appendix:Agent Prompt Details}

\definecolor{backcolour}{rgb}{0.95,0.95,0.92}

\lstdefinelanguage{json}{
    basicstyle=\normalfont\ttfamily,
    showstringspaces=false,
    breaklines=true,
    backgroundcolor=\color{backcolour},
    commentstyle=\color{gray}\ttfamily,
    keywordstyle=\color{blue}\ttfamily,
    stringstyle=\color{blue}\ttfamily,
    keywordstyle={[2]\color{blue}\ttfamily},
    morekeywords={[2] 
        \{\%, \%\}, \{\{, \}\},
    },
    literate=
        {\{\%}{{\textcolor{blue}{\{\%}}}{1}
        {\%\}}{{\textcolor{blue}{\%\}}}}{1}
        {\{\{}{{\textcolor{blue}{\{\{~}}}{1}
        {\}\}}{{\textcolor{blue}{~\}\}}}}{1},
    sensitive=true,
    frame=single, %
    framesep=1em, %
    xleftmargin=1em, %
    xrightmargin=1em, 
    morekeywords={video_height,video_width,task_prompt,sub_task_list, current_task, advice_,  drag_start_width, drag_start_height, drag_end_width, drag_end_height, center_width, center_height, operation_type,  operation_value, is_last_action_in_subsession, website, domain, subdomain },
}

Here are the prompts for different roles. For certain datasets, there are some changes in their data format.\\

\textbf{Prompt for HumanEval, MBPP} \\

Planning phase prompt template:
\begin{lstlisting}[language=json]
prompt = (
          f"Provide guided steps to solve the following problem and identify potential challenges.: {question}. "
          f"[requirement]: less text, don't give code"
)
\end{lstlisting}

Coding phase prompt template:
\begin{lstlisting}[language=json]
prompt = (
          f"As a code expert, according to the guidance:{design_solution}"
          f"please provide a python solution to the following programming problem: {question}."
          f"Ensure that the answer produced by your code matches the test cases in the examples:{test_case}"
          f"[Important]only give the code and should not include any explanations or comments. "
)
\end{lstlisting}

Debugging phase prompt template:
\begin{lstlisting}[language=json]
prompt = (
          f"According to the {question}, the code given is:{implementation_solution} "
          f":Fix it using traceback:{result_traceback}. "
          f"[Important]Only give code don't analyze and no annotation"
)
\end{lstlisting}

Super-Role's prompt template:
\begin{lstlisting}[language=json]
prompt = (
          f"According to the problem:{question}"
          f"Use the experience to give the code to solve it, make sure it will pass the text case:{test_case}"

)
\end{lstlisting}

\vspace{60pt}
Super-Role's refinement template:
\begin{lstlisting}[language=json]
prompt = (
          f"For this problem, {question}, your previous answer encountered an error: {first_solution}. "
          f"Traceback: {result}. "
          f"To proceed, ensure the new solution meets the following requirements:\n"
          f"1. Is fundamentally different from the previous solution.\n"
          f"2. Fixes the above error.\n"
          f"3. Passes all the given test cases: {test_case}.\n\n"
          f"Here are some examples: {Example}. "
          f"Hint: Try to explore different logic or structures, such as using loops, functions, or list comprehensions.\n\n"
)
\end{lstlisting}

\textbf{Prompt for APPS} \\

Planning phase prompt template:
\begin{lstlisting}[language=json]
prompt = (
          f"Provide guided steps to solve the following problem and identify potential challenges.: {question}. "
          f"[requirement]: less text, don't give code"
)
\end{lstlisting}

Coding phase prompt template:
\begin{lstlisting}[language=json]
prompt = (
          f"As a code expert, according to the guidance:{design_solution}"
          f"please provide a python solution to the following programming problem: {question}."
          f"Ensure that the answer produced by your code matches the test cases in the examples:{test_case}"
          f"The function name must be the same as in the problem{prompt_name}" 
          f"[Important]only give the code and should not include any explanations or comments. "
)
\end{lstlisting}

Debugging phase prompt template:
\begin{lstlisting}[language=json]
prompt = (
          f"According to the {question}, the code given is:{implementation_solution} "
          f":Fix it using traceback:{result_traceback}. "
          f"[Important]Only give code don't analyze and no annotation"
          f"Make sure the function name is the same as in the problem{prompt_name}"

)
\end{lstlisting}

\vspace{60pt}

Super-Role's prompt template:
\begin{lstlisting}[language=json]
prompt = (
          f"According to the problem:{question}"
          f"Use the experience to give the code to solve it, make sure it will pass the text case:{test_case}"
          # f"Use the same function name in the problem{prompt_name}"
          f"[Important]:Only codes. No comments or annotation"
)
\end{lstlisting}

Super-Role's refinement template:
\begin{lstlisting}[language=json]
prompt = (
          f"For this problem, {question}, your previous answer encountered an error: {first_solution}. "
          f"Traceback: {result}. "
          f"To proceed, ensure the new solution meets the following requirements:\n"
          f"1. Is fundamentally different from the previous solution.\n"
          f"2. Fixes the above error.\n"
          f"3. Passes all the given test cases: {test_case}.\n\n"
          f"Here are some examples: {output}. "
          f"Hint: Try to explore different logic or structures, such as using loops, functions, or list comprehensions.\n\n"
          f"[requirement]: Only codes. No comments or annotation"
          f"Use the same function name in the problem{prompt_name}"
)
\end{lstlisting}

\textbf{Prompt for xCodeEval, CodeContest} \\
Due to our use of a unified test, we have forced both the input and output to be a single string parameter. While this approach standardizes the operation, it does not guarantee 100\% success in defining the function, which can lead to discrepancies between the test results and reality. We recommend integrating a standard test and removing the forced content to achieve better results.

Planning phase prompt template:
\begin{lstlisting}[language=json]
prompt = (
          f"Provide guided steps to solve the following problem and identify potential challenges.: {question}. "
          f"[requirement]: less text, don't give code"
)
\end{lstlisting}

Coding phase prompt template:
\begin{lstlisting}[language=json]
prompt = (
          f"As a code expert, according to the guidance:{design_solution}"
          f"please provide a python solution to the following programming problem: {question}."
          f"Ensure that the answer produced by your code matches the test cases in the examples:{test_case}"
          f"[Important]only give the code and should not include any explanations or comments. "
          f"[Important]:Use a function to solve the problem, ending with a return.All the code is inside the function."
          f"Make sure the function only requires a single string parameter."
)
\end{lstlisting}

Debugging phase prompt template:
\begin{lstlisting}[language=json]
prompt = (
          f"According to the {question}, the code given is:{implementation_solution} "
          f":Fix it using traceback:{result_traceback}."
          f"[Important]Only give code don't analyze and no annotation"
          f"[Important]:Use a function to solve the problem, ending with a return."
          f"Make sure the function only requires a single string parameter.All the code is inside the function."
          f"Only code no comments or other things"
)
\end{lstlisting}

Super-Role's prompt template:
\begin{lstlisting}[language=json]
prompt = (
          f"According to the problem:{question}"
          f"Use the experience to give the code to solve it, make sure it will pass the text case:{test_case}"
          f"[Important]:Only codes. No comments or annotation"
          f"Use a function to solve the problem, ending with a return, and only require a single string parameter"
          f"All the code is inside the function."
)
\end{lstlisting}

Super-Role's refinement template:
\begin{lstlisting}[language=json]
prompt = (
          f"For this problem, {question}, your previous answer encountered an error: {first_solution}. "
          f"Traceback: {result}. "
          f"To proceed, ensure the new solution meets the following requirements:\n"
          f"1. Is fundamentally different from the previous solution.\n"
          f"2. Fixes the above error.\n"
          f"3. Passes all the given test cases: {test_case}.\n\n"
          f"Here are some examples: {output}. "
          f"Hint: Try to explore different logic or structures, such as using loops, functions, or list comprehensions.\n\n"
          f"[requirement]: Only codes. Make only require a single string parameter"
          f"All the code is inside the function."
          f"code only require a single string parameter"  #codecontest
)
\end{lstlisting}

\begin{figure*}[]
    \centering
    \includegraphics[width=1\linewidth]{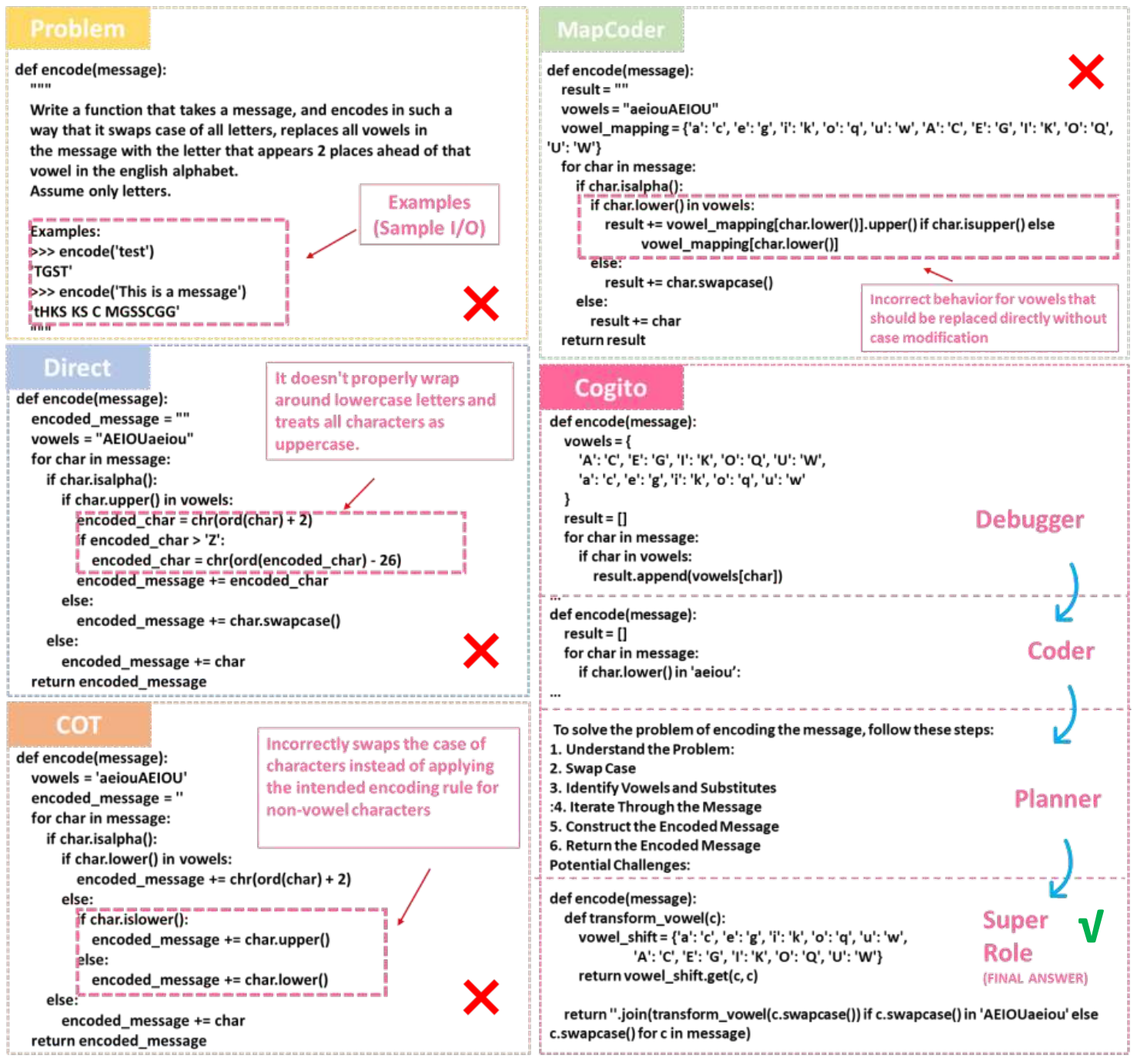}
    \caption{An example answer for 5 methods of HumanEval \#92.}
    \label{fig:appendix_case}
\end{figure*}

\twocolumn

\section{Supplementary Experimental Details}

\subsection{Datasets}
\label{appendix:datasets}

For convenience, we used the HumanEval dataset from Mapcoder~\cite{2024_ACL_MapCoder}, which contains a sample column that separately extracts the execution examples provided in the prompt, making it easier to execute and return results. Similarly, in MBPP, they also select some data from the test set as inputs, but maintain the independence of the test set and the exclusivity between MBPP and MBPP-ET. For CodeContest, we only use the test section consisting of 165 problems. APPS and xCodeEval utilize a subset of problems extracted from the raw data by MapCoder.

\noindent\subsection{Evaluation Metric}
\label{appendix:evaluation-metric}

The unbiased version of \textbf{Pass @k} is a metric commonly used to evaluate the effectiveness of recommendation systems, especially when the system's recommendations are prone to biases, such as those arising from uneven distributions of relevant items or recommendation behaviors. This metric aims to provide a more equitable and robust evaluation by correcting for these biases, which can distort performance assessments.

The formula for \textbf{Pass@k} is given by:

\begin{equation}
\text{Pass@k} = \mathbb{E}_{\text{Problems}} \left[ 1 - \frac{\binom{n-c}{k}}{\binom{n}{k}} \right],
\end{equation}

Here, \( n \) represents the total number of items available for recommendation, \( c \) stands for the number of relevant items (those that the user is interested in), and \( k \) denotes the size of the top-k recommendations generated by the system. The metric aims to assess the probability that at least one relevant item appears within the top-k recommended items.

The expression \( \frac{\binom{n-c}{k}}{\binom{n}{k}} \) calculates the probability that none of the relevant items appear in the top-k recommendations. It does so by computing the number of ways to choose \( k \) items from the \( n-c \) non-relevant items, normalized by the total number of ways to choose \( k \) items from all \( n \) items. Subtracting this probability from 1 yields the probability that at least one relevant item is present in the top-k recommendations.

By considering combinations, the \textbf{Pass@k} metric effectively accounts for the distribution of relevant items and corrects for any inherent biases in the recommendation process. This is particularly important because traditional metrics might overestimate the performance of systems that tend to recommend a small subset of highly popular items, leaving out less obvious but still relevant items. The unbiased formula ensures that the evaluation is fairer, reflecting a more realistic measure of a system’s ability to recommend relevant items across a diverse set of users or test cases.

The expectation \( \mathbb{E}_{\text{Problems}} \) averages the performance over a set of users or test cases, ensuring that the metric evaluates the system’s performance across different scenarios rather than focusing on a single instance. This provides a more comprehensive measure of the system’s overall effectiveness, as it considers the variability of user preferences and the diversity of possible recommendation outcomes.

In practical terms, \textbf{Pass@k} is useful for recommendation systems where the goal is to not only recommend relevant items but to ensure that these recommendations are evenly distributed across all relevant possibilities, avoiding any bias toward popular items or particular types of content. This makes it a more robust and fair metric for evaluating real-world recommendation engines.

\noindent\subsection{Baselines}
\label{appendix:baselines}
We evaluate our approach by comparing it with several baseline methods. First, we use the \textbf{Direct} Method, where the prompt is submitted to the LLM without decomposition to assess its intrinsic reasoning. We then evaluate two structured reasoning methods: Chain-of-Thought (\textbf{CoT}), which solves the problem step-by-step, and \textbf{Self-Planning}, which separates planning and implementation phases. Our approach, which incorporates GitHub searches for relevant code, is compared with \textbf{Analogical Reasoning}, a retrieval-based method. Finally, we include \textbf{Mapcoder}, a state-of-the-art method, as a benchmark. All tests are conducted using GPT-3.5-turbo (GPT-3.5-turbo-0125) and GPT-4 (GPT-4-0613) from OpenAI.

\subsection{An example answer for 5 methods of HumanEval \#92}
We present solutions from several methods for the 92nd problem in HumanEval (Figure~\ref{fig:appendix_case}). Except for Cogito, all fail the first test: "assert candidate('TEST') == 'tgst', 'This prints if this assert fails 1 (good for debugging!)'". Each method’s errors are highlighted, with explanations provided. For \textbf{Cogito}, we show its responses at different stages.

\subsection{The comparison between GPT-3.5-TURBO and GPT-4 responses.}
We select two examples to demonstrate the performance differences between GPT-3.5 and GPT-4. The first example, from the HumanEval dataset (Figure~\ref{HumanEval_case}), illustrates a case where GPT-3.5 fails to produce a correct solution, while GPT-4 successfully generates the correct code. Similarly, the second example from the APPS dataset (Figure~\ref{APPS_case}) highlights a scenario where GPT-3.5 struggles, but GPT-4 is able to handle the problem effectively. Upon examining both cases, it becomes clear that GPT-4 excels in solving more complex and challenging problems, showing improved reasoning and code generation capabilities compared to GPT-3.5.

\subsection{The complete response process of Cogito in APPS \#1628}
We provide the responses from Cogito and the roles involved throughout the entire process tackling Task 1628 of APPS (Figure~\ref{fig:appendix_case_complete}). The responses of adjacent roles in each group are also provided, which help to give a better context of cause and effect. By showcasing the interactions between roles, we can better understand how different stages of the process contribute to the final outcome.Additionally, we present the errors, changes, and correct parts in each response, breaking down the reasons behind each modification. This allows us to explain why the code either passes or fails, highlighting specific areas where improvements are made or where missteps occurr.





\begin{figure*}[]
    \centering
    \includegraphics[width=\textwidth]{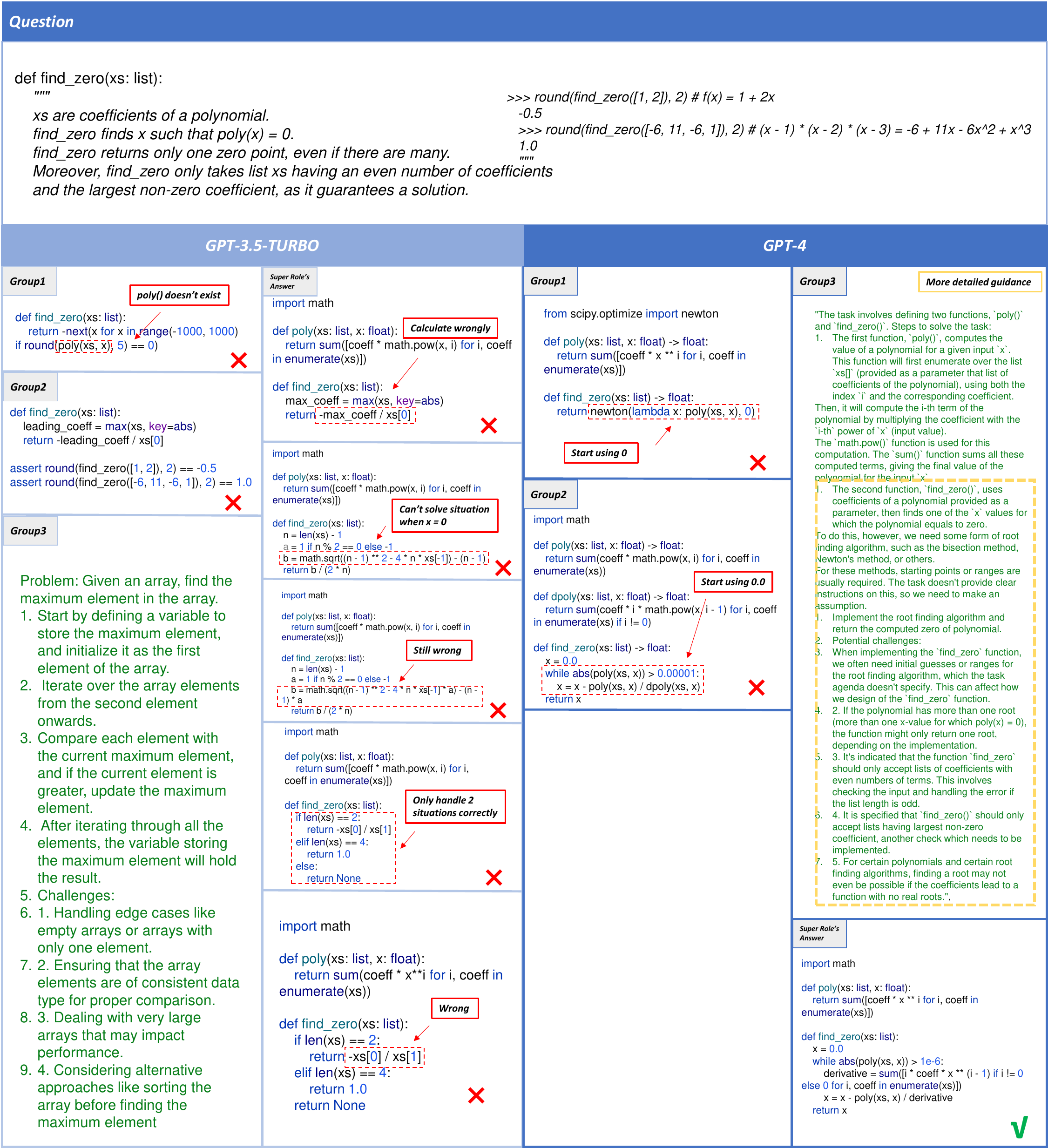}
    \caption{The complete solution process of Cogito in APPS\#1628.}
    \label{fig:appendix_case_complete}
\end{figure*}

\begin{figure*}[]
    \centering
    \includegraphics[width=\textwidth]{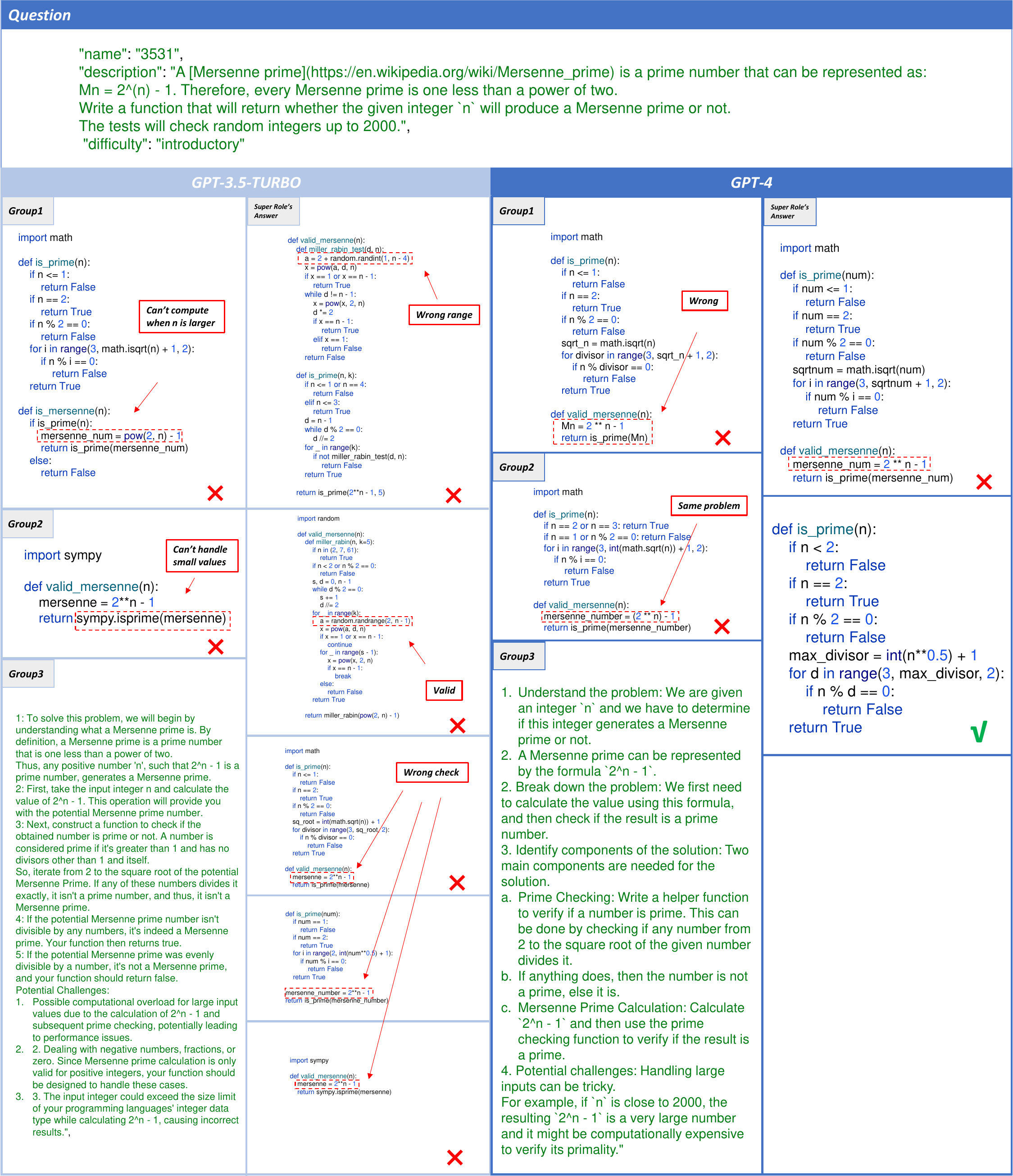}
    \caption{The complete solution process of Cogito in HumanEval\#92.}
    \label{fig:appendix_case_complete}
\end{figure*}

\begin{figure*}[]
    \centering
    \includegraphics[width=\textwidth]{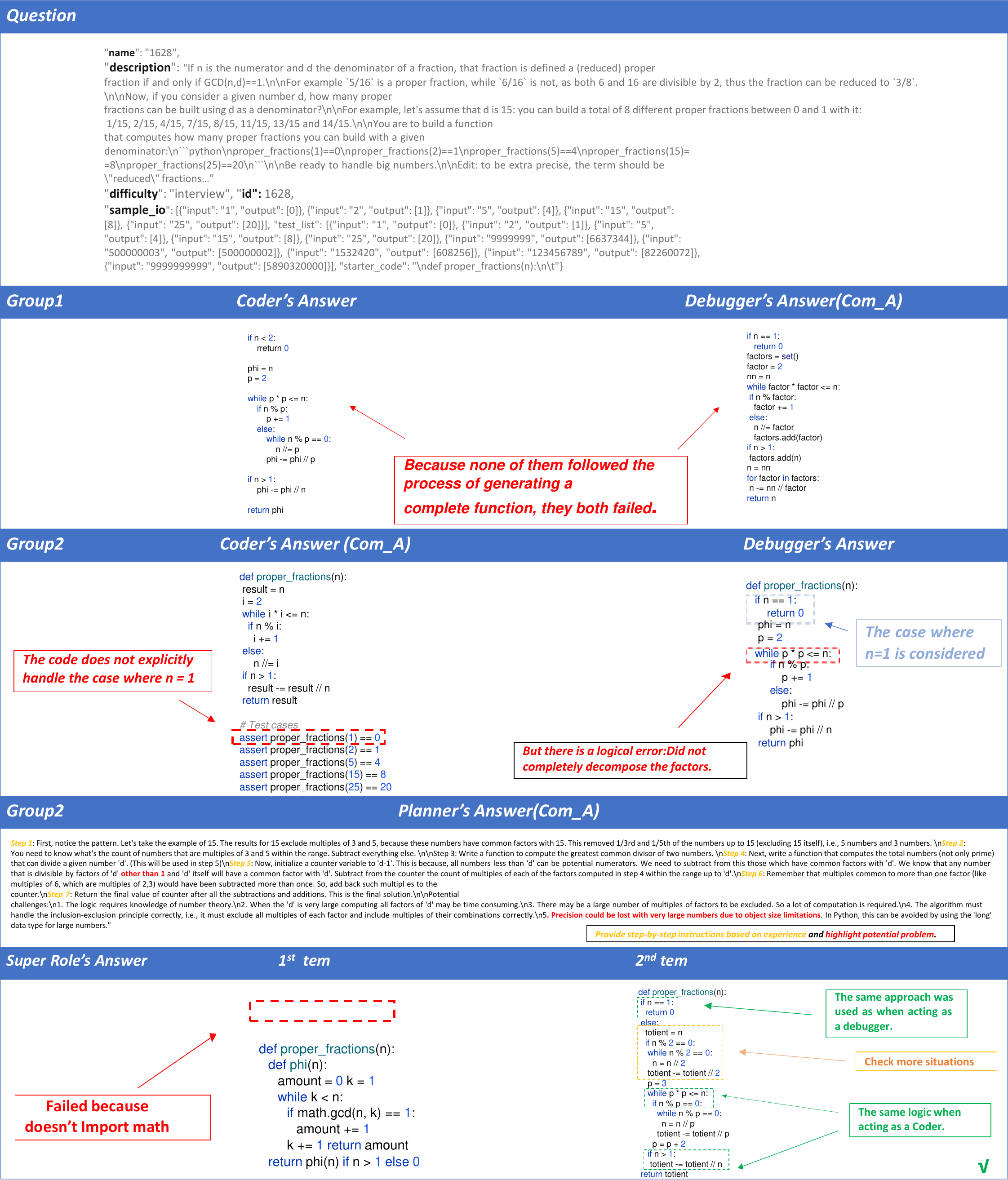}
    \caption{The complete solution process of Cogito in APPS\#1628.}
    \label{fig:appendix_case_complete}
\end{figure*}

\end{document}